\documentclass[hyper,preprint]{JHEP3} 
\JHEPspecialurl{http://jhep.sissa.it/JOURNAL/JHEP3.tar.gz}
\usepackage{epsfig,multicol}

\def\o{\omega}
\def\tr{\mbox{Tr}}
\def\be{\begin{equation}}
\def\ee{\end{equation}}
\def\bea{\begin{eqnarray}}
\def\eea{\end{eqnarray}}

\title{The Leigh-Strassler Deformation and the Quest for Integrability}
\author{T. M\aa nsson\\
        Max-Planck Institut f\"ur Gravitationsphysik,\\
Albert-Einstein-Institut
 Am M\"uhlenberg 1, D-14476 Potsdam, Germany\\
        E-mail:  \email{teresia@aei.mpg.de}}
\preprint{\hepth{0703150}\\AEI-2007-011}

\abstract{In this paper we study the one-loop dilatation operator of the full
scalar field sector of Leigh-Strassler deformed $\mathcal{N}$=4 SYM theory. In particular we
map it onto a spin chain and find the parameter values for which the Reshetikhin
integrability criteria are fulfilled.  
Some years ago Roiban found an integrable subsector, consisting of two holomorphic
scalar fields, corresponding to the $XXZ$ model. He was pondering about the existence of
a subsector which would form generalisation of that model to an integrable $\mathfrak{su}_q(3)$ model.
Later Berenstein and Cherkis added one more holomorphic field and showed that the subsector
obtained this way cannot be integrable except for the case when $q=e^{i\beta}$, $\beta \in \mathcal{R}$.
In this work we show if we add an anti-holomorphic field to the two holomorphic
ones, we get indeed an integrable $\mathfrak{su}_q(3)$ subsector.
We find it plausible that a direct generalisation to a $\mathfrak{su}_q(2|3)$ one-loop 
sector will exist, and possibly beyond one-loop.}

\begin{document}

\section{Introduction}
In  the past years, Maldacena's AdS/CFT correspondence \cite{Maldacena:1998re,Gubser:1998bc,Witten:1998qj}
has received much attention due to its great potential to solve non-perturbative problems.
Later it was discovered that the dilatation operator
for the SYM theory can be mapped to a spin chain Hamiltonian, which turns out
to be integrable \cite{Minahan:2002ve,Beisert:2003yb,Beisert:2003tq}.
This has greatly simplified the quest for a proof or disproof of the conjecture.
Significant progress have been made lately to find a full string S-matrix to be mapped to
an asymptotic all loop gauge theory S-matrix in order to prove the
 correspondence \cite{Beisert:2006ez,Beisert:2006ib}.

In order to come closer to  more realistic models people have tried
to extend the duality for different deformations. One natural
starting point is to understand the deformations which preserve
the conformal invariance of the theory. 
There is a particular family of deformations, parametrized by two complex numbers $q$ and $h$, which preserve both the conformal symmetry and one of the supersymmetries, called the Leigh-Strassler deformation
 \cite{Leigh:1995ep}. Actually finiteness to all orders have
only been proven for $q=e^{i\beta}$, $\beta$ real\cite{Ananth:2006ac,Rossi:2006mu,Elmetti:2006gr}. 
One question that has arisen is if there is a connection between
integrability and finiteness. We will exhibit a non-trivial example,
where the integrability condition and the condition for two-loop finiteness agree
perfectly.

On the supergravity side  a way of generating supergravity
duals to the $\beta$-deformed field theory was introduced in \cite{Lunin:2005jy},
 and in \cite{Frolov:2005dj} it was used to construct a
three-parameter generalization of the $\beta$-deformed theory. 
There also have been  some attempts to construct
backgrounds for non-zero $h$ 
\cite{Aharony:2002hx,Fayyazuddin:2002vh,Kulaxizi:2006zc}.
 In \cite{Niarchos:2002fc} 
the BMN limit for the theory was considered. 

In \cite{Frolov:2005ty,Frolov:2005iq,Kuzenko:2005gy} agreement
 between the supergravity sigma model and the coherent state action
coming from the spin chain describing the $\beta$-deformed
dilatation operator was shown. 
The
gauge theory dual to the three parameter supergravity deformation was found in
\cite{Frolov:2005dj,Beisert:2005if} for $q_j=e^{i\gamma_j}$ with $\gamma_j$
real, corresponding to certain phase deformations in the Lagrangian.
The
$\beta$-deformed theory is obtained when all the $\gamma_j=\beta$. The
result is that the theory is integrable for
 any $q=e^{i\gamma_j}$ with $\gamma_j$ real \cite{Beisert:2005if}. The
general case with complex $\gamma_j$ is not integrable
\cite{Berenstein:2004ys,Freyhult:2005ws}.
The authors in \cite{Alday:2005ww}  developed a general procedure to obtain
the string Green-Schwarz action, and in particular they derived the 
monodromy matrix for the $\gamma$-deformations on the string side.

 Integrability, on the gauge theory side, has been 
investigated in a number of papers \cite{Roiban:2003dw,
Berenstein:2004ys,Beisert:2005if,Freyhult:2005ws,Bundzik:2005zg}.
First Roiban  \cite{Roiban:2003dw} discovered that the one-loop dilatation
 operator
in the holomorphic two-field subsector
corresponds to the integrable XXZ-spin chain. He also discuss the possibility
that this result might generalise to a $\mathfrak{su}_q(3)$ sector. 
Then Berenstein and Cherkis \cite{Berenstein:2004ys} showed that integrability
is only preserved for special values of $q$ ($q=e^{i\beta}$ where $\beta \in \mathcal{R}$)
 when one more holomorphic field is included. Here we find 
that if you instead add a non-holomorphic field to the theory, you get
a closed sector which is indeed an integrable $\mathfrak{su}_q(3)$ sector.
Integrable Hamiltonians of this form were classified in \cite{Freyhult:2005ws}. 
We notice that our Hamiltonian just differs (besides some phases) from the
usual $\mathfrak{su}_q(3)$ model, often called the trigonometric (or hyperbolic) $A_{2}$ vertex model
 \cite{DeVega:1988ry}, with an additional term that cancel  for periodic spin chains.
 This is also
another example when the condition on the prefactor in front of the F-term, required
 for the theory to be integrable, and
the finiteness condition coincide.  It would therefore be very interesting
if the integrable sector can be extended to the  $\mathfrak{su}_q(2|3)$ and then to all loops.
Higher loop generalisations have been studied in the context of the Hubbard model for
$\mathfrak{su}_{\gamma}(2)$, $\mathfrak{sl}_{\gamma}(2)$ and  $\gamma \in \mathcal{R}$ \cite{McLoughlin:2006cg},
and for $GL$ models in \cite{Beisert:2005wv}.

In another related work \cite{Bundzik:2005zg} the spin-chain obtained with both $q$ and $h$
non-zero was considered. There a set of integrable
values for $h$ and $q$ was found, and it was also suggested that maybe an 
elliptic R-matrix could give rise to more cases.
In this work we will extend the analysis of the last paper to show that
the elliptic R-matrix of Belavin \cite{Belavin:1981ix}, which has the right symmetries to give rise to
the Hamiltonian, gives rise only in exceptional cases to Hermitian 
matrices, and not to any more cases than the ones already found.
We will use the Reshetikhin's criteria for integrability to discard
the possibility of finding any more integrable cases than the ones 
found in \cite{Bundzik:2005zg} obtained from R-matrices of trigonometric or
elliptic types.

The analysis will be extended to include the full  one loop scalar field
sector of the theory.
We conclude that all integrable cases, but those corresponding
to diagonal Hamiltonians in the holomorphic sector, also satisfy the
Reshetikhin's condition in the full scalar field sector. We also notice
that the relations between the $q$-deformed case with $h=0$ 
and $h$-deformed case with $q=0$  gets destroyed in the full
sector. In the end we will compare the spectra of the two cases.

An outline of the paper is as follows.
We will first start with reviewing the Reshetikhin's criteria for integrability
in section two. 
In the third section the non-holomorphic sector will be analysed. In
section four the three spin sector with $\mathfrak{su}_q(3)$ symmetry will be presented.
Finally in section five we show for the holomorphic sector that no more cases
can be obtained from the Reshetikhin condition. Finally in Appendix (\ref{Appendix2}) we add a discussion about the
Belavin R-matrix.

\section{The Reshetikhin condition}
An integrable, nearest neighbour interaction, spin chain Hamiltonian
can be obtained from an R-matrix as the first charge:
\be
Q_1=H=\sum_{i=1}^L \,P\partial_u R_{i,i+1}|_{u=0}=\sum_{i=1}^L h_{i,i+1}.
\label{forsta laddning}
\ee 
The R-matrix satisfies the Yang-Baxter equation:
\begin{equation}
 R_{12}(u-v)R_{13}(u)R_{23}(v)=R_{23}(v)R_{13}(u)R_{12}(u-v),
\end{equation}
where we have assumed that $R$ is of genus one or less, so that
its dependence on the spectral parameters $u$ and $v$ can always be
put into the form $u-v$ \cite{Gomez:1996az}. Therefore the following is only valid for
trigonometric and elliptic, but not hyper-elliptic R-matrices, {\it e.g.} as in
the chiral Potts model. Instead of working with the R-matrix above, we could
work with the permuted version, defined as $\hat{R}=PR$, where $P$ is the
permutation matrix. Depending on the situation, we will find it more convenient
to refer to $R$ or $\hat{R}$ as the R-matrix. We can always choose to rescale
the R-matrix, so that $\hat{R}(0)=1$.

By use of the unitarity condition, $\hat{R}(u)\hat{R}(-u)=I\otimes I$, the
second charge can be shown to be
\be
Q_2=i\sum_{i=1}^L[h_{i,i+1},h_{i+1,i+2}]\,.
\ee
Conditions on the complex parameters $q$ and $h$ in the Hamiltonian are obtained
by demanding that this charge commutes with $H$. With the existence of the boost 
operator it is easy to understand that integrability then follows. Here we will
give a short explanation of this. Tetelman \cite{Tetelman:1982} showed that all the commuting charges
$Q_n$ of an integrable spin chain, with the above first and second charge, can
be generated iteratively as
\be
Q_{n+1}=[B,Q_n]\;,
\ee
where $B$ is the boost operator \cite{Tetelman:1982}, which is defined as
\be
B=\sum k h_{k,k+1}\;.
\ee
In particular, the $Q_2$ charge above can be generated in this way from $Q_1$.
It is then easy to prove that $[Q_1,Q_2]=0$ implies $[Q_1,Q_n]=0$ 
for all integer $n>1$. This in turn implies that all the $Q_n$
commute. But note that the boost operator is formally only defined for infinitely
long spin chains ($L\rightarrow \infty$).

The commutator is calculated to be
\bea
&[Q_1,Q_2]&=\sum_{i=1}^L h_{i,i+1}^2\,h_{i+1,i+2}-
h_{i,i+1}h_{i,i+2}^2+h_{i+1,i+2}^2h_{i,i+1}-
h_{i+1,i+2}h_{i,i+1}^2\\
&&-2\,h_{i,i+1}h_{i+1,i+2}h_{i,i+1}+2\,h_{i+1,i+2}h_{i,i+1}h_{i+1,i+2}\nonumber
\eea
This vanishes whenever the Reshetikhin's condition
\be
[h_{i,i+1}+h_{i+1,i+2},[h_{i,i+1},h_{i+1,i+2}]]=I\otimes A-A\otimes I,
\ee
is satisfied. In the following section we will find for which parameter 
values the Reshetikhin's condition holds, for the full scalar field one-loop dilatation
operator. In the last section we will check that the Reshetikhin's criteria
in the holomorphic sector is only fulfilled for the intergable cases
already known.

\section{The non-holomorphic sector}
Next we will extend the analysis of reference \cite{Bundzik:2005zg} to include the full
scalar field sector, in accordance with \cite{Minahan:2002ve} for the non-deformed
theory. When the dilatation operator acts on both holomorphic and anti-holomorphic fields,
the cyclicity of the trace will give rise to contributions from rotating the
interactions (see the illustration below).
There is also an extra contribution from the D-term. It is because of this piece that,
quite remarkably, the non-deformed case is still integrable in the full scalar field
sector. Luckily, the diagrams coming from photon interactions do not alter things to first
loop order, and the fermion contribution to the self energy is the same as in the
non-deformed case, and will only give contributions proportional to the identity matrix.
The D-term scalar field contribution is
\be
\mathcal{L}_D=\frac{g^2}{2}\tr[\phi_i,\bar{\phi}_i][\phi_j,\bar{\phi}_j]
=\frac{g^2}{2}
\tr\left(
\phi_i\bar{\phi}_i\phi_j\bar{\phi}_j+ \bar{\phi}_i\phi_i\bar{\phi}_j\phi_j
-\phi_i\bar{\phi}_i\bar{\phi}_j\phi_j-\bar{\phi}_i \phi_i\phi_j\bar{\phi}_j
\right)
\ee 
where a summation over $i,j=0,1,2$ is understood. The indices of the fields
$\phi_i$ are identified modulo three. The action of the dilatation operator on a general
operator $O=\psi^{i_1\ldots i_L}\tr\,\phi_{i_1}\ldots \phi_{i_L}$, to first loop order,
can be deduced using Feynman graphs and a regularisation in accordance with
\cite{Roiban:2003dw,Berenstein:2004ys}.

The trace operators, $O=\psi^{i_1\ldots i_L}\tr\,\phi_{i_1}\ldots \phi_{i_L}$,
will be mapped to spin chain states
$|\Psi\rangle=\psi^{i_1\ldots i_L}| i_1\ldots i_L\rangle $.
The action of the dilatation operator is translated into actions on spin chain states,
which can be illustrated graphically as

\begin{picture}(70,100)
\put(20,30){\line(1,1){40}}
\put(20,70){\line(1,-1){40}}
\put(6,20){$|1\rangle$}
\put(6,75){$|2\rangle$}
\put(63,20){$|2\rangle$}
\put(63,75){$|1\rangle$}
\put(18,0){$\phi_2\phi_1 \bar{\phi}_2\bar{\phi}_1$}
\put(120,30){\line(1,1){40}}
\put(120,70){\line(1,-1){40}}
\put(106,20){$|2\rangle$}
\put(106,75){$|\bar{1}\rangle$}
\put(163,20){$|\bar{1}\rangle$}
\put(163,75){$|2\rangle$}
\put(118,0){$\bar{\phi}_1 \phi_2\phi_1\bar{\phi}_2$}
\put(220,30){\line(1,1){40}}
\put(220,70){\line(1,-1){40}}
\put(206,20){$|\bar{1}\rangle$}
\put(206,75){$|\bar{2}\rangle$}
\put(263,20){$|\bar{2}\rangle$}
\put(263,75){$|\bar{1}\rangle$}
\put(218,0){$\bar{\phi}_2\bar{\phi}_1\phi_2\phi_1$}
\put(320,30){\line(1,1){40}}
\put(320,70){\line(1,-1){40}}
\put(306,20){$|\bar{2}\rangle$}
\put(306,75){$|1\rangle$}
\put(363,20){$|1\rangle$}
\put(363,75){$|\bar{2}\rangle$}
\put(318,0){$\phi_1\bar{\phi}_2\bar{\phi}_1 \phi_2$}
\end{picture}

{\flushleft All} the graphs in this example originate from the action of
 the same trace term
in the Lagrangian, $\tr\,\phi_2\phi_1\bar{\phi}_2\bar{\phi}_1$.

By introducing the operators $E_{ij}$, which act on the basis states as
$E_{ij}|k\rangle=\delta_{jk}|i\rangle$, the dilatation operator can be written as a
spin-chain Hamiltonian with nearest-neighbour interactions, {\it i.e.}
$\Delta\propto \sum_l (h^D_{l,l+1}+h^F_{l,l+1}+I)$, where $H^D$ is the D-term contribution and
$H^F$ is the F-term contribution and the contribution to the identity matrix, $I$, comes from the
one-loop self energy diagram and the boson exchange diagram. The D-term part of the Hamiltonian is
\bea
h_{l,l+1}^D = && +  E_{\bar{i}\bar{j}}\otimes E_{i j}+ E_{ij}\otimes E_{\bar{i}\bar{j}}
+E_{\bar{j}\bar{j}}\otimes E_{ii}+E_{ii}\otimes E_{\bar{j}\bar{j}} \qquad i\neq j\\
&&+2\,E_{\bar{i}\bar{i}}\otimes E_{ii}+2\,E_{ii}\otimes E_{\bar{i}\bar{i}}
-E_{\bar{i}i}\otimes E_{i\bar{i}}-E_{\bar{j}i}\otimes E_{j\bar{i}} \\
&&-E_{j \bar{i}}\otimes E_{\bar{j} i}-E_{i i}\otimes E_{i i}
-E_{\bar{i} \bar{i}}\otimes E_{\bar{i} \bar{i}}
- E_{ii}\otimes E_{jj}+ E_{\bar{i}\bar{i}}\otimes E_{\bar{j}\bar{j}}
-E_{i \bar{i}}\otimes E_{\bar{i} i}
\eea
Likewise, we can write down the F-term part of the scalar field Lagrangian
\footnote{As mentioned in \cite{Madhu:2007ew,Freedman:2005cg} there should also be some double trace contributions for the $SU(N)$ gauge group,
but they go as $1/N$ and therefore only affect the anomalous dimension of operators involving two scalar fields.}
\bea
 \mathcal{L}_F &=& \frac{4g^2}{1+q^*q+h^*h}\tr\left[\phi_i
\phi_{i+1}\bar{\phi}_{i+1}\bar{\phi}_i
 -q\phi_{i+1}\phi_i\bar{\phi}_{i+1}\bar{\phi}_i
 - q^{*} \phi_i\phi_{i+1}\bar{\phi}_i\bar{\phi}_{i+1} \right]\nonumber \\
&+&\tr\left[ qq^{*}\phi_{i+1}\phi_i\bar{\phi}_i\bar{\phi}_{i+1}
-qh^{*} \phi_{i+1}\phi_i \bar{\phi}_{i+2}\bar{\phi}_{i+2}
-q^{*}h \phi_{i+2}\phi_{i+2}\bar{\phi_i}\bar{\phi}_{i+1}\right]\nonumber \\
&+&\tr\left[ h\phi_{i+2}\phi_{i+2}\bar{\phi}_{i+1}\bar{\phi}_i
+h^{*} \phi_i\phi_{i+1}\bar{\phi}_{i+2}\bar{\phi}_{i+2} +hh^{*}
\phi_{i}\phi_{i}\bar{\phi}_i\bar{\phi}_i \right], 
\eea
and the F-term part of the Hamiltonian
\bea
 h_{l,l+1}^F &=& \frac{4}{1+q^*q+h^*h}\left(
 E_{i,i}\otimes E_{i+1,i+1}  -q
E_{i+1,i}\otimes E_{i,i+1}  -q^{*}
E_{i,i+1}\otimes E_{i+1,i}\right. \nonumber \\ 
&+&qq^{*} E_{i+1,i+1}\otimes   E_{i,i}  -qh^{*}
 E_{i+1,i+2}\otimes E_{i,i+2}   -q^{*}h
E_{i+2,i+1}\otimes E_{i+2,i}  \nonumber \\
  &+&  hE_{i+2,i}\otimes E_{i+2,i+1}
+h^{*}E_{i,i+2}\otimes E_{i+1,i+2} +hh^{*}E_{i,i}\otimes
E_{i,i} \nonumber 
\\
&+& E_{\bar{i},i+1}\otimes E_{i,\overline{i+1}} 
-q E_{\bar{i},i+1}\otimes E_{i+1,\bar{i}}  
-q^{*} E_{\overline{i+1},i}\otimes E_{i,\overline{i+1}} \nonumber \\
&+&qq^{*} E_{\overline{i+1},i}\otimes  E_{i+1,\bar{i}} 
 -qh^{*}
 E_{\overline{i+2},i+2}\otimes E_{i+1,\bar{i}}  
 -q^{*}h
E_{\overline{i+1},i}\otimes E_{i+2,\overline{i+2}}  \nonumber \\ 
 &+&  h E_{\bar{i},i+1}\otimes E_{i+2,\overline{i+2}}
+h^{*}E_{\overline{i+2},i+2}\otimes E_{i,\overline{i+1}} 
+hh^{*}E_{\bar{i},i}\otimes E_{i,\bar{i}} 
\\
&+& E_{\overline{i+1},\overline{i+1}}\otimes E_{\bar{i},\bar{i}} 
 -q
E_{\overline{i+1},\bar{i}}\otimes E_{\bar{i},\overline{i+1}}  
-q^{*}
E_{\bar{i},\overline{i+1}}\otimes E_{\overline{i+1},\bar{i}}
 \nonumber \\ 
&+&qq^{*} E_{\bar{i},\bar{i}}\otimes  E_{\overline{i+1},\overline{i+1}}
  -qh^{*}
 E_{\overline{i+2},\bar{i}}\otimes E_{\overline{i+2},\overline{i+1}} 
  -q^{*}h
E_{\bar{i},\overline{i+2}}\otimes E_{\overline{i+1},\overline{i+2}}
  \nonumber \\
  &+&  hE_{\overline{i+1},\overline{i+2}}\otimes E_{\bar{i},\overline{i+2}}
+h^{*}E_{\overline{i+2},\overline{i+1}}\otimes E_{\overline{i+2},\bar{i}}
 +hh^{*}E_{\bar{i},\bar{i}}\otimes
E_{\bar{i},\bar{i}} \nonumber 
\\
&+& E_{i+1,\overline{i}}\otimes E_{\overline{i+1},i} 
 -q
E_{i,\overline{i+1}}\otimes E_{\bar{i+1},i}  
-q^{*}
E_{i+1,\overline{i}}\otimes E_{\overline{i},i+1}
 \nonumber \\ 
&+&qq^{*} E_{i,\overline{i+1}}\otimes  E_{\overline{i},i+1}
  -qh^{*}
 E_{i,\overline{i+1}}\otimes E_{\overline{i+2},i+2} 
  -q^{*}h
E_{i+2,\overline{i+2}}\otimes E_{\overline{i},i+1}
  \nonumber \\
  &+&  hE_{i+2,\overline{i+2}}\otimes E_{\overline{i+1},i}
+h^{*}E_{i+1,\overline{i}}\otimes E_{\overline{i+2},i+2}
 +hh^{*}E_{i,\bar{i}}\otimes
E_{\bar{i},i}\,, \nonumber 
\label{spin-chain-Hamiltonian1}
\eea
where the coefficient in front, $4/(1+q^*q+h^*h)$, comes from the
two-loop finiteness condition \cite{Parkes:1984dh,Jones:1984cx}.

Now, we like to go through all parameter values for which the holomorphic sector is
integrable, and check if Reshetikhin's condition is still satisfied. That is, whether the
matrix
$$\mathcal{DU}={[h_{l,l+1}+h_{l+1,l+2},[h_{l,l+1},h_{l+1,l+2}]]}$$
can, for the different values of the parameters, be written in the form
\be
\label{requested form}
A\otimes I-I\otimes A.\,.
\ee

But first, let us consider whether there are some cases that can immediately
be understood to be integrable, besides the ones related by a local transformation,
$$(U\otimes U) h_{l,l+1}\,(U\otimes U)^{-1},$$ 
to the $q$-deformed case with $q=e^{i\beta}$
and $h=0$, for real $\beta$. When $q=(1+\rho)e^{i2\pi m/3}$ and
$h=\rho e^{i2\pi n/3}$, and $m$ and $n$ integers, the phases can be transformed away, as
in the holomorphic sector. It means that this last case is related to $q=e^{i\beta}$
and $h=0$ via the local
transformation mentioned, plus a site dependent phase shift.

The question we may ask then is whether the $q$-deformed and $h$-deformed cases are related
in the full scalar field sector, via the same type of non-local transformation
as for the holomorphic sector \cite{Bundzik:2005zg}. This transformation
is site dependent and acts on a spin state $|a\rangle_l$, where $l$ is the site number, as
 \be \label{ekv:trans3} |a\rangle_{1+3k}\rightarrow
|a-1\rangle_{1+3k}\,, \;\;\;\; |a\rangle_{2+3k}\rightarrow
|a+1\rangle_{2+3k}\,, \;\;\;\; |a\rangle_{3k}\rightarrow
|a\rangle_{3k}\,, \ee where $a$ takes the values $0$, $1$ or $2$.
The anti-holomorphic sector will transform  opposite to the one above:
\be \label{ekv:trans4} |\bar{a}\rangle_{1+3k}\rightarrow
|\overline{a+1}\rangle_{1+3k}\,, \;\;\;\; |\bar{a}\rangle_{2+3k}\rightarrow
|\overline{a-1}\rangle_{2+3k}\,, \;\;\;\; |\bar{a}\rangle_{3k}\rightarrow
|\bar{a}\rangle_{3k}\,, \ee

Below we illustrate graphically how the interaction terms coming from
$\tr\,\phi_2\phi_2\bar{\phi_0}\bar{\phi_{1}}$ transform under the action of
(\ref{ekv:trans3}, \ref{ekv:trans4}).
In the diagram, it is assumed that the leftmost site is $3k$ and the rightmost site is
$3k+1$. Therefore, the transformation does not change the leftmost spin states in the
diagram below:

\begin{picture}(70,190)
\put(20,110){\line(1,1){40}}
\put(20,150){\line(1,-1){40}}
\put(6,100){$|1\rangle$}
\put(6,155){$|2\rangle$}
\put(63,100){$|0\rangle$}
\put(63,155){$|2\rangle$}
\put(120,110){\line(1,1){40}}
\put(120,150){\line(1,-1){40}}
\put(106,100){$|0\rangle$}
\put(106,155){$|\bar{1}\rangle$}
\put(163,100){$|\bar{2}\rangle$}
\put(163,155){$|2\rangle$}
\put(220,110){\line(1,1){40}}
\put(220,150){\line(1,-1){40}}
\put(206,100){$|\bar{2}\rangle$}
\put(206,155){$|\bar{0}\rangle$}
\put(263,100){$|\bar{2}\rangle$}
\put(263,155){$|\bar{1}\rangle$}
\put(320,110){\line(1,1){40}}
\put(320,150){\line(1,-1){40}}
\put(306,100){$|\bar{2}\rangle$}
\put(306,155){$|2\rangle$}
\put(363,100){$|1\rangle$}
\put(363,155){$|\bar{0}\rangle$}
\put(20,20){\line(1,1){40}}
\put(20,60){\line(1,-1){40}}
\put(6,10){$|1\rangle$}
\put(6,65){$|2\rangle$}
\put(63,10){$|2\rangle$}
\put(63,65){$|1\rangle$}
\put(120,20){\line(1,1){40}}
\put(120,60){\line(1,-1){40}}
\put(106,10){$|0\rangle$}
\put(106,65){$|\bar{1}\rangle$}
\put(163,10){$|\bar{0}\rangle$}
\put(163,65){$|1\rangle$}
\put(220,20){\line(1,1){40}}
\put(220,60){\line(1,-1){40}}
\put(206,10){$|\bar{2}\rangle$}
\put(206,65){$|\bar{0}\rangle$}
\put(263,10){$|\bar{0}\rangle$}
\put(263,65){$|\bar{2}\rangle$}
\put(320,20){\line(1,1){40}}
\put(320,60){\line(1,-1){40}}
\put(306,10){$|\bar{2}\rangle$}
\put(306,65){$|2\rangle$}
\put(363,10){$|0\rangle$}
\put(363,65){$|\bar{1}\rangle$}
\put(37,80){$\Downarrow$}
\put(137,80){$\Downarrow$}
\put(237,80){$\Downarrow$}
\put(337,80){$\Downarrow$}
\end{picture}

{\flushleft We} see that the first three interaction terms, which we get from transforming the $h$-deformed
case, exist in the $q$-deformed Hamiltonian. However the fourth
term is not an interaction which exists for the $q$-deformed case. From this the conclusion is that this
transformation cannot relate the $q$-deformed and the $h$-deformed case in the non-holomorphic sector. In
figure (\ref{fig:spektra1}) we can see how the energy eigenvalues differ between the $q$-deformed and $h$-deformed cases. 
There are similarities, but also some significant differences.
\FIGURE[t]{\label{fig:spektra1}
\parbox{7cm}{\centering\includegraphics[height=6cm]{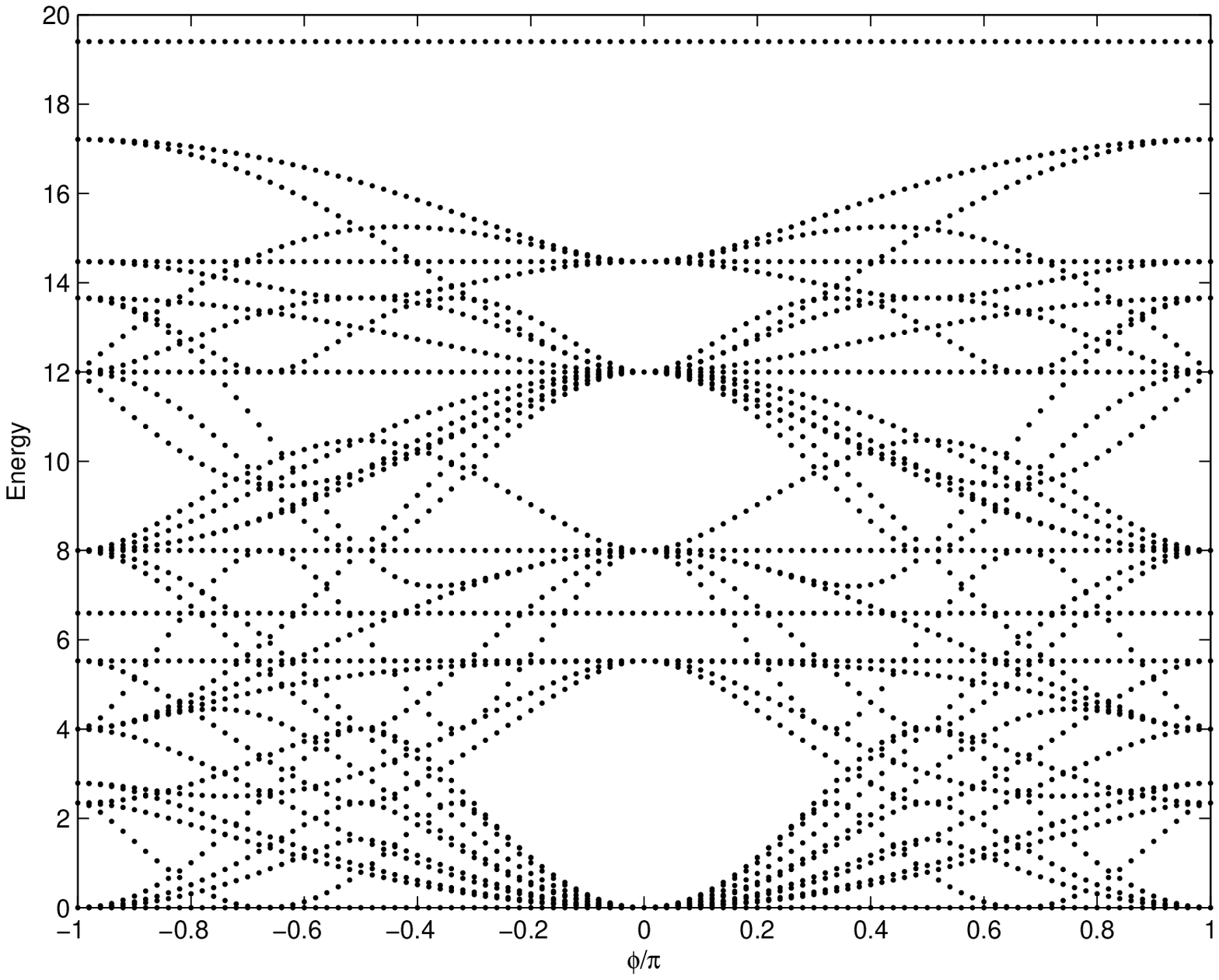}}
$\;\;$
\parbox{7cm}{\centering\includegraphics[height=6cm]{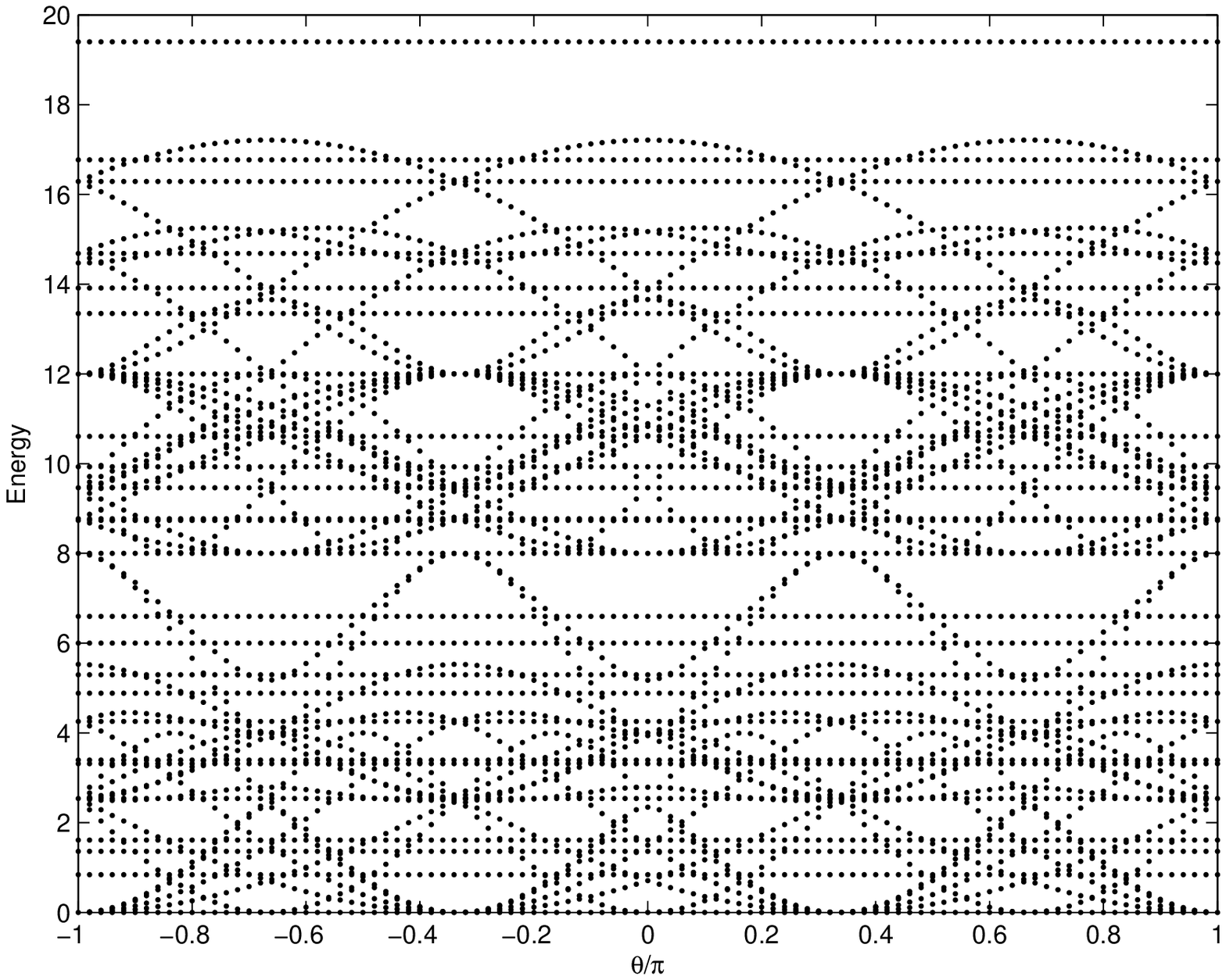}}
\caption{Spin chain with four sites. The left graph shows the
energy spectrum as a function of the phase $\phi$, when
 $q=e^{i \phi}$ and $h=0$. The right graph shows the
spectrum as a function of the phase $\theta$, when
$h=e^{i\,\theta}$ and $q=0$. } }

Now we are ready to start examining for which values of the parameters 
that Reshetikhin's condition is still satisfied in the non-holomorphic sector.
By simply looking at the matrix $\mathcal{DU}$, we see that this cannot
be the case when both $h$ and $q$ vanish, and also when $q=-1$, $h=1$.
At a first glance, the rest of the cases seem very promising. To ensure that Reshetikhin's condition
is satisfied, we do the following.
The matrix element
$$\mathcal{DU}_{(36(j_1-1)+6(n-1)+k_1,36(j_2-1)+6(m-1)+k_2)}$$
with $j_1\neq j_2$ and $k_1\neq k_2$ needs to vanish. It corresponds to
the matrix element 
$$\alpha_{j_1j_2nmk_1k_2}E_{j_1j_2}\otimes E_{nm}\otimes E_{k_1k_2}$$
Here we have chosen to rename the indices, so that barred ones correspond to odd values and non-barred correspond to even values of the original indices.
The result with Mathematica is that all these terms vanish. It can be checked that if $k_1\neq k_2$,
then the following holds:
$$\alpha_{jjnmk_1k_2}=-\alpha_{nmk_1k_2\tilde{j}\tilde{j}}$$
where $j$ and $\tilde{j}$ take any values. Thus, all these terms
satisfy the requested form (\ref{requested form}). We also see that for
$n\neq m$ the same holds true:
$$\alpha_{jjnmkk}=-\alpha_{nm kk\tilde{j}\tilde{j}}\, ,$$
again for any $j$ and $\tilde{j}$.
The only thing that remains to be checked is whether the coefficients, $\alpha_{jjnnkk}$, come 
out the right way. To simplify our notation, we define $\tilde{\alpha}_{jnk}\equiv\alpha_{jjnnkk}$.
For a consistency check we can do the case $q=e^{i\beta}$ first, which we know is integrable. In
this case, the non-zero matrix elements are

\begin{tabular}{lllll}
 $\tilde{\alpha}_{2k,2k,2k-1}$&$=-22$, & &$\tilde{\alpha}_{2k-1,2k,2k}$&$=22$\\
 $\tilde{\alpha}_{2k-1,2k-1,2k}$&$=-22$,& &$\tilde{\alpha}_{2k,2k-1,2k-1}$&$=22$, \\
$\tilde{\alpha}_{n,2k-1,2k}$&$=-6$, & &$\tilde{\alpha}_{2k,2k-1,n}$&=$6$, \\
$\tilde{\alpha}_{n,2k,2k-1}$&$=-6$, & &$\tilde{\alpha}_{2k-1,2k,n}$&$=6$, \\
$\tilde{\alpha}_{n,2k,2k}$&$=16$, & &$\tilde{\alpha}_{2k,2k,n}$&$=-16$, \\
$\tilde{\alpha}_{n,2k-1,2k-1}$& $=16$,& & $\tilde{\alpha}_{2k-1,2k-1,n}$ &$=-16$,\\
\end{tabular}

{\flushleft where} $n\neq 2k,2k-1$. In order to see that the above
is of the requested form (\ref{requested form}),
the $\tilde{\alpha}_{j_1,j_2,j_3}$ above can be devided into a part
 $\tilde{\alpha}^R_{j_1,j_2,j_3}$ coming
from the term of the form $I\otimes A$
and a part
 $\tilde{\alpha}^L_{j_1,j_2,j_3}$ coming
from  $A\otimes I$. Thus 
$\tilde{\alpha}_{j_1,j_2,j_3}=\tilde{\alpha}^R_{j_1,j_2,j_3}+\tilde{\alpha}^L_{j_1,j_2,j_3}$ with:

\begin{tabular}{lll}
$\tilde{\alpha}^R_{l,2k-1,2k}=-6$,&& $\tilde{\alpha}^L_{2k,2k-1,l}=6$,\\
$\tilde{\alpha}^R_{l,2k,2k-1}=-6$,&& $\tilde{\alpha}^L_{2k-1,2k,l}=6$,\\
$\tilde{\alpha}^R_{l,2k,2k}=16$,&& $\tilde{\alpha}^L_{2k,2k,l}=-16$,\\
$\tilde{\alpha}^R_{l,2k-1,2k-1}=16$,&& $\tilde{\alpha}^L_{2k-1,2k-1,l}=-16$,\\
\end{tabular}

{\flushleft here $l$ takes any value}.
Thus the terms are of the right form (\ref{requested form}), as
they should.
Now we do the same thing for the case $q=0$ and $h=e^{i\beta}$. 
In this case we cannot take the integrability for granted, since we have
 not found
any transformation relating it to the former case, as in the holomorphic 
sector.
If we repeat the above analysis, we obtain

\begin{tabular}{lllll}
 $\tilde{\alpha}_{2k+3,2k+4,2k}$&$=22$, & &$\tilde{\alpha}_{2k+1,2k+5,2k}$&$=-22$\\
 $\tilde{\alpha}_{2k+2,2k+1,2k+5}$&$=22$,& &$\tilde{\alpha}_{2k+4,2k,2k+1}$&$=-22$, \\
$\tilde{\alpha}_{n_1,2k+1,2k+5}$&$=16$, & &$\tilde{\alpha}_{2k+1,2k+5,n_2}$&=$-16$, \\
$\tilde{\alpha}_{n_1,2k+4,2k}$&$=16$, & &$\tilde{\alpha}_{2k+4,2k,n_{3}}$&$=-16$, \\
$\tilde{\alpha}_{2k+5,2k,n_3}$& $=6$,& & $\tilde{\alpha}_{n_{4},2k+5,2k}$ &$=-6$,\\
$\tilde{\alpha}_{2k,2k+5,n_{2}}$&$=6$, & &$\tilde{\alpha}_{n_5,2k,2k+5}$&$=-6$, \\
\end{tabular}

{\flushleft where} $n_1\neq 2k+2,2k+3$, $n_2\neq 2k+3,2k$, 
$n_3\neq 2k+5,2k+2$, $n_{4}\neq 2k,2k+1$, $n_5\neq 2k+4,2k+5$. The terms above can be organized in the following
way, which turns out also to be 
of the required form (\ref{requested form}) 

\begin{tabular}{lll}
$\tilde{\alpha}^R_{l,2k+1,2k+5}=16$,&& $\tilde{\alpha}^L_{2k+1,2k+5,l}=-16$,\\
$\tilde{\alpha}^R_{l,2k+4,2k}=16$,&& $\tilde{\alpha}^L_{2k+4,2k,l}=-16$,\\
$\tilde{\alpha}^R_{2k+5,2k,l}=6$,&& $\tilde{\alpha}^L_{l,2k+5,2k}=-6$,\\
$\tilde{\alpha}^R_{2k,2k+5,l}=6$,&& $\tilde{\alpha}^L_{l,2k,2k+5}=-6$.\\
\end{tabular}

{\flushleft Thus} we conclude that also the $h$-deformed case with $h=e^{i\beta}$ and $q=0$
is integrable. 

An interesting thing is to note  that the case $q=-1$ and $h=1$ (which is
equivalent to the $h\rightarrow \infty$ case) would
have remained integrable, if it were not for the extra contribution
of the D-term. Without the D-term contribution, the Hamiltonian turns out to be
a sum of three decoupled Heisenberg spin chains.
But even if the full Hamiltonian in these cases no longer fulfill
Reshetikhin's condition, we can find some subsectors where the Hamiltonian is
diagonalizable. To identify these subsectors, we start by analysing
just the D-term.

\subsection*{The D-term subsectors}
We will start by looking for integrable subsectors of the D-term. We have seen
that the full D-term does not satisfy the Reshetikhin's condition, but 
 it indeed consists of several subsectors where the Hamiltonian
can immediately be diagonalised. One thing we notice at once is that there is a 
subsector where the eigenstates are of the form
\be
|a\,b\,a\,b\,a\,b\,a\,b\rangle, \qquad a=1 \, \mbox{or} \, \bar{1} \qquad
 b=2 \, \mbox{or} \, \bar{2}
\ee
Acting on a state like this with the Hamiltonian simply gives
\be
(m-n+L)|a\,b\,a\,b\,a\,b\,a\,b\rangle, \qquad n\in Z, \qquad m\in Z,
\ee
where $n$ is the number of all states of the type $|\bar{k}\,\bar{l} \rangle$ or
$|k\,l\rangle$, and $m$ is the number of all states of the type $|k\,\bar{l} \rangle$ or
$|\bar{k}\,l\rangle$, and $L$ is the total number of states $|a\rangle$, $|b\rangle$. 
We see that if we add the F-term contribution when
$h\rightarrow\infty$ and also when both $q=0$ and $h=0$, these states
will still be eigenstates. 

Another diagonal subsector can be built up of states consisting of
\bea
&&|A\rangle=|3\rangle \otimes (|2\rangle\otimes|\bar{2}\rangle-|1\rangle\otimes |\bar{1}\rangle)\otimes |3\rangle \\
&&|B\rangle=|3\rangle \otimes (|\bar{2}\rangle \otimes |2\rangle-|\bar{1}\rangle \otimes |1\rangle)\otimes |3\rangle
\eea
acting with the Hamiltonian on $|A\rangle$, respectively $|B\rangle$, gives
\be
H|A\rangle=4|A\rangle, \qquad H|B\rangle=4|B\rangle, \qquad \mbox{and} 
\qquad H\,|A\rangle \otimes |B\rangle=10 |A\rangle \otimes |B\rangle
\ee
We can continue making this little game also when adding F-terms coming from the 
diagonal ones in the holomorphic sector. Here we will do it for the $q=0$
and $h=0$ case. If we make an Ansatz that this  state will be diagonal,
\be
|Zup\rangle=|3\rangle\otimes (\alpha \,|2\rangle \otimes |\bar{2}\rangle+\beta \,|1\rangle \otimes |\bar{1}\rangle +
\gamma \,|\bar{2}\rangle\otimes |2\rangle+\delta \,|\bar{1}\rangle\otimes |1\rangle)|3\rangle
\ee 
then it will be an eigenstate if $\gamma=\beta=0$ and $\alpha=\delta$. We will call this
eigenstate $|Zip\rangle$,
\be
 |Zip\rangle=|3\rangle\otimes (\alpha \,|2\rangle \otimes |\bar{2}\rangle+\delta \,|\bar{1}\rangle\otimes |1\rangle)|3\rangle\,.
\ee
With the Hamiltonian acting on the open state $|Zip\rangle$
\be
H|Zip\rangle=8|Zip\rangle 
\ee
It will also be  an  eigenstate if  $\gamma=-\beta$ and $\delta=-\alpha$ and $Y_{\pm}=\alpha/\beta=-2\pm \sqrt{5}$, 
namely
\be
 |Zap_{\pm}\rangle=|3\rangle\otimes (Y_{\pm} \,|2\rangle \otimes |\bar{2}\rangle+ |1\rangle \otimes |\bar{1}\rangle -
|\bar{2}\rangle\otimes |2\rangle- Y_{\pm}\,|\bar{1}\rangle\otimes |1\rangle)|3\rangle\,.
\ee
The Hamiltonian acting on the open state $|Zap\rangle$ gives
\be
H|Zap_{\pm}\rangle=(10+2Y_{\pm})|Zap_{\pm}\rangle 
\ee
From this more general eigenstates with combinations of $Zip$ and $Zap$ states can be constructed,
\be
|Zip\rangle^{n_1}|Zap_{\pm}\rangle^{n_2} \ldots|Zip\rangle^{n_{45}}\,.
\ee
\section{An integrable sector with $\mathfrak{su}_q (3)$ symmetry}
In this section we will show the existence of an integrable three-state subsector of the full scalar
field theory, with $\mathfrak{su}_q (3)$ symmetry, for general complex values for $q$, with $h=0$.
We notice, from the interaction terms in the full Hamiltonian (\ref{spin-chain-Hamiltonian1}), that
such a subsector exists. It consists of the states $|a\rangle$ with $a=i,\overline{i+1},i+2$,
{\it e.g.} $a=1,\bar{2},3$. This subsector has two holomorphic and one anti-holomorphic field (not a
conjugate of any of the two holomorphic ones). The nearest neigbour part of the 
Hamiltonian in this sector is
\bea
&&h_{i,i+1}=\frac{2}{|q|^{-1}+|q|}\left(\frac{|q|^{-1}+|q|}{2}\left(E_{11}\otimes E_{\bar{2}\bar{2}}+ E_{\bar{2}\bar{2}}\otimes E_{11}+E_{00}\otimes E_{\bar{2}\bar{2}}+ E_{\bar{2}\bar{2}}\otimes E_{00}\right)+ \right.\nonumber \\
&&+\left(|q|^{-1}E_{11}E_{00}+|q|\,E_{00}\otimes E_{11}\right) \nonumber \\
&&\left.
-\left(e^{i\beta}\,(E_{10}\otimes E_{01}+E_{1\bar{2}}\otimes E_{\bar{2}1}+E_{\bar{2}0}\otimes E_{0\bar{2}})+h.c.\right)
\right) \nonumber
\label{Hamiltonian:SUq3}
\eea
The phase $\beta$ is defined by $q=|q|e^{i\beta}$. We notice in passing that when $q=1$, this
becomes the ordinary Heisenberg spin-chain Hamiltonian.
In \cite{Freyhult:2005ws}, all integrable Hamiltonians with symmetry $U(1)^3$ were classified. 
It is easily seen that the Hamiltonian above has this symmetry. 
First of all, it was shown that phases can be transformed away,
as they did not affect the Yang-Baxter equation. Therefore the phase $e^{i\beta}$ can be disregarded
from the further analysis. We now set it equal to one, and when we later write  the R-matrix for the
system, we can re-insert the phases if we so desire.

This implies that we can immediately check the integrability conditions
 given in
\cite{Freyhult:2005ws}. But first, let us introduce some notation.
 The Hamiltonian is written as
$$h_{i,i+1}=h_{kn}^{lm}\,E_{lk}\otimes E_{mn}.$$ The norm of the off-diagonal elements are all equal, and 
denoted by $r$.
 Then the integrability condition, which was obtained in 
 \cite{Freyhult:2005ws} from demanding that the S-matrix satisfying
the Yang-Baxter equation, reads \footnote{Reshetikhin's condition
leads to somewhat stronger constraints because it does not allow for
the freedom a Hamiltonian with an $U(1)^3$ symmetry has to
 add number operators ({\it i.e.} of the form $E_{11}\otimes I$) which
commute with the Hamiltonian. This freedom needs to be added by hand.
But for the Hamiltonian (\ref{Hamiltonian:SUq3}) it can
be immediately checked that it satisfies Reshetikhin's condition.}
\be
\tau_1\,\tau_2=r^2 , \qquad \sigma_1=\sigma_2=\tau_1+\tau_2,
\ee
where
\bea
&&\tau_1=h_{10}^{10}-h_{00}^{00}-h_{12}^{12}+h_{02}^{02}=
\frac{2|q|^{-1}}{|q|^{-1}+|q|},\\
&&\tau_1=h_{01}^{01}-h_{00}^{00}-h_{21}^{21}+h_{20}^{20}=
\frac{2|q|}{|q|^{-1}+|q|}, \\
&&
\sigma_1=h_{10}^{10}-h_{00}^{00}-h_{11}^{11}+h_{01}^{01}=2, \\
&&
\sigma_2=h_{20}^{20}-h_{00}^{00}-h_{22}^{22}+h_{02}^{02}=2.
\eea
In our case,
\be
r=\frac{2}{|q|^{-1}+|q|}.
\ee
Thus the Hamiltonian satisfies the integrability conditions. We see that the condition
on the term in front of the F-term obtained from integrability is the same as the one coming
from the finiteness condition.
In \cite{Freyhult:2005ws}, the
Hamiltonians with six off-diagonal elements were divided in three classes. We see that our Hamiltonian
 belongs to
the same class as the Hamiltonian with $\mathfrak{su}_q(3)$ symmetry mentioned in \cite{Berenstein:2004ys}, which 
came from the $A_{2}$ hyperbolic R-matrix \cite{DeVega:1988ry}. Actually after a closer look we notice
that these Hamiltonians are in fact  related to each other by a term 
$$
\frac{|q|^{-1}-|q|}{|q|^{-1}+|q|}\left(E_{\bar{2}\bar{2}}\otimes I-I\otimes E_{\bar{2}\bar{2}}\right).
$$
This term will cancel out for a periodic spin chain like the one we have,
 but even so
we might want to find
the explicit R-matrix for our Hamiltonian (\ref{Hamiltonian:SUq3}). Actually, it turns out to be very
easy in this case, because we have two
different types of subsectors which are related to each other in a very neat way. The first is
spanned by $|1\rangle$ and $|\bar{2}\rangle$, or $|0\rangle$ and $|\bar{2}\rangle$, and the second by
$|1\rangle$ and $|0\rangle$. The Hamiltonian of the latter can be written, up to a multiplicative
factor $2/(|q|^{-1}+|q|)$, as
\be
h_{i,i+1}=e_i=\pmatrix{
  0 & 0 & 0 & 0  \cr
  0 & |q|^{-1} &  -1 & 0 \cr
  0 & -1 & |q| & 0 \cr
  0 & 0 & 0 & 0
  }.
\label{Hamiltonian:temperley}
\ee
The latter is a nice (and famous) Hamiltonian. It satisfies the Temperley-Lieb-Jones algebra
\bea
&&e_ie_{i+1}e_i=e_i \\
&&e_ie_j=e_je_i \qquad;\qquad |j-i| >2 \\
&&e_i^2=(|q|+|q|^{-1}).
\eea
which makes it possible to write the $R$-matrix for this in a very nice form \cite{Gomez:1996az}
\be
\hat{R}=I+\frac{\sinh\, (u)}{\sinh\,(\gamma-u)}\,e_i, \qquad |q|=e^{\gamma}.
\label{lilla R-matris}
\ee
This R-matrix gives the Hamiltonian (\ref{Hamiltonian:temperley}), up to a multiplicative factor
$1/(|q|-|q|^{-1})$, through the procedure (\ref{forsta laddning}).

Moving on to the Hamiltonian for the subsector spanned by $|1\rangle$ and $|\bar{2}\rangle$, or
$|0\rangle$ and $|\bar{2}\rangle$, is:
\be
h_{i,i+1}=\pmatrix{
  0 & 0 & 0 & 0  \cr
  0 & \frac{|q|+|q|^{-1}}{2} &  -1 & 0 \cr
  0 & -1 & \frac{|q|+|q|^{-1}}{2} & 0 \cr
  0 & 0 & 0 & 0
  }.
\label{Hamiltonian:XXZ}
\ee
This one has the form of the usual XXZ two-particle $R$-matrix:
\be
\hat{R}(u)=\pmatrix{
  1 & 0 & 0 & 0  \cr
  0 & b &  c & 0 \cr
  0 & c & b & 0 \cr
  0 & 0 & 0 & 1
  }, \qquad b=\frac{\sinh(\gamma)}{\sinh(\gamma-u)},\qquad c=\frac{\sinh(u)}{\sinh(\gamma-u)},
\ee
with $|q|=e^{\gamma}$. We see that the $c$ is the same as in the other $R$-matrix
(\ref{lilla R-matris}). The derivative of $c$ taken at zero is $1/\sinh(\gamma)=1/(|q|-|q|^{-1})$,
and the derivative of $b$ at zero is  $\cosh(\gamma)/\sinh(\gamma)=(|q|+|q|^{-1})/(|q|-|q|^{-1})$.
Thus we get the Hamiltonian (\ref{Hamiltonian:XXZ}) up to the same multiplicative factor as when we got 
the Hamiltonian (\ref{Hamiltonian:temperley}) case when
extracting the Hamiltonian through (\ref{forsta laddning}).

Matching these two R-matrices together we then have to check that the terms in the Yang-Baxter
equation involving all three fields cancel each other out. We have verified that this is the case, and we
get a total R-matrix for the full Hamiltonian (\ref{Hamiltonian:SUq3})
\bea
\hat{R}(u)=&&E_{11}E_{11}+E_{\bar{2}\bar{2}}E_{\bar{2}\bar{2}}+E_{33}E_{33}+\\
&&
+\frac{\sinh(\gamma)}{\sinh(\gamma-u)}\left(E_{11}E_{\bar{2}\bar{2}}+ E_{\bar{2}\bar{2}}E_{11}+E_{33}E_{\bar{2}\bar{2}}+ E_{\bar{2}\bar{2}}E_{33}\right)\\
&&
(1+\frac{\sinh(u)\,e^{\gamma}}{\sinh(\gamma-u)})E_{11}E_{33}+(1+\frac{\sinh(u)\,e^{-\gamma}}{\sinh(\gamma-u)})E_{33}E_{11} \\
&&
-\frac{\sinh(u)}{\sinh(\gamma-u)}\left(e^{i\beta}\,(E_{13}E_{31}+E_{1\bar{2}}E_{\bar{2}1}+E_{\bar{2}3}E_{3\bar{2}})+h.c.\right)
\eea
This R-matrix just gives rise to the same Bethe equations as the $\mathfrak{su}_q(3)$ invariant $A_2$ model when using
the nested Bethe Ansatz, if $\beta$ is zero \cite{DeVega:1988ry}. This is because the only difference between this R-matrix
and the R-matrix for the $A_2$ model is the terms $\hat{R}_{ij}^{ij}$, $i\neq j$. These terms cancels out of the
Bethe equations because 
$$\frac{\hat{R}_{ij}^{ij}(u)}{\hat{R}_{ij}^{ji}(u)}=-\frac{\hat{R}_{ij}^{ij}(-u)}{\hat{R}_{ij}^{ji}(-u)}.$$
In \cite{Beisert:2005if} the authors studying R-matrices with phases appearing in the same way as in our
R-matrix, and we can use the result given
there to include the phases.
After including these phases we get the following set of
algebraic Bethe equations: 
\begin{equation}
 \prod_{l\neq k}^{K_2} \frac{\sinh((\mu_{2,k}-\mu_{2,l})-\gamma)}{\sinh((\mu_{2,k}-\mu_{2,l})+\gamma)}\prod_{j=1}^{K_1} 
\frac{\sinh((\mu_{2,k}-\mu_{1,j})+\gamma/2)}{\sinh ((\mu_{2,k}-\mu_{1,j})-\gamma/2)}=
e^{i(2\,n_1+n_2+n_3)\beta}
\end{equation}
\bea
 &&e^{-i(n_1+n_2)\beta}\left(
\frac{\sinh (\mu_{1,k}+\gamma/2)}{\sinh(\mu_{1,k}-\gamma/2)}\right)^L=\\ 
&&\prod_{l\neq k}^{K_1} \frac{\sinh((\mu_{1,k}-\mu_{1,l})+\gamma)}{\sinh((\mu_{1,k}-\mu_{1,l})-\gamma)}\,
\prod_{j=1}^{K_2} \frac{\sinh((\mu_{1,k}-\mu_{2,j})-\gamma/2)}{\sinh((\mu_{1,k}-\mu_{2,j})+\gamma/2)}
\eea
as well as the cyclicity constraint:
\begin{equation}
 e^{i(-n_2+n_3)\beta}\prod_{l}^{K_1} \frac{\sinh(\mu_{1,l}+\gamma/2)}{\sinh(\mu_{1,l}-\gamma/2)}=1
\end{equation}
Here we have chosen a notation where by  $n_1$ is related to $|\bar{2}\rangle$, $n_2$ is related to $|3\rangle$
and $n_3$ is related to $|1\rangle$. Also, $L=n_1+n_2+n_3$ and $K_2=n_2+n_3$.

\section{The holomorphic sector}
The one-loop dilatation operator for the Leigh-Strassler deformation in the
holomorphic sector was given in \cite{Bundzik:2005zg} as a nearest neighbour
spin chain Hamiltonian. Now we are interested in whether this spin chain 
Hamiltonian is integrable for any other parameter values than in the cited work.
 There, the authors hoped for the existence of an R-matrix which
would give rise to more cases. Indeed, an elliptic R-matrix exists, the Belavin
R-matrix \cite{Belavin:1981ix}, which has the right symmetry. 
But as we will  see, unfortunately neither it, nor any
other elliptic or trigonometric R-matrix, can possibly give rise to any more cases. 

We
start out by writing out the above mentioned spin chain Hamiltonian such that
all explicit interactions are clearly visible:
\be
H=\sum_l h_{l,l+1},
\ee
where
\bea
\label{ekv:dil}
h_{l,l+1} &=& E_{i,i}\otimes E_{i+1,i+1}  -q
E_{i+1,i}\otimes E_{i,i+1}  -q^{*}
E_{i,i+1}\otimes E_{i+1,i} \nonumber \\
&+&qq^{*} E_{i+1,i+1}\otimes   E_{i,i}  -qh^{*}
 E_{i+1,i+2}\otimes E_{i,i+2}   -q^{*}h
E_{i+2,i+1}\otimes E_{i+2,i}  \nonumber
\\&+&hE_{i+2,i}\otimes E_{i+2,i+1}
+h^{*}E_{i,i+2}\otimes E_{i+1,i+2} +hh^{*}E_{i,i}\otimes
E_{i,i}\, . \label{spin-chain-Hamiltonian} 
\eea 
Here, the operators $E_{ij}$ are defined to act on the basis states as
$E_{ij}|k\rangle=\delta_{jk}|i\rangle$. The Hamiltonian can be rewritten in a
form which makes the $Z_3\times Z_3$ symmetry more apparent:
\be
\label{ekv:Hamiltonian1}
h_{l,l+1}=\sum_{n,m=0}^2
\,\o_{nm} S_{n,m}\otimes S_{2n,2m} \, .
\ee
The generators $S_{n,m}$ can be defined in terms of a product 
$S_{n,m}=e^{i2\pi nm/3}S^n_{1,0}S^m_{0,1}$ with
\be
S_{1,0}=\pmatrix{
  0 & 1 & 0   \cr
  0 & 0 &  1 \cr
  1 & 0 & 0   }, \qquad 
S_{0,1}=\pmatrix{
  e^{\frac{-i2\pi}{3}} & 0 & 0   \cr
  0 &1 & 0\cr
  0 & 0 &  e^{\frac{i2\pi}{3}}
  }.
\ee
The generators are related by
\bea
S_{n,m}S_{l,k} &=& e^{-i\frac{2\pi (nk-ml)}{3}}S_{n+m,k+l}\,,
\eea
where the indices are defined modulo three.
The coefficients $\o_{nm}$ are related to $h$ and $q$ as
\bea
&& \o_{0,k}=\left(h\,h^*+q\,q^*\,e^{-\frac{i\,2\,\pi\,k}{3}}+
e^{\frac{i\,2\,\pi\,k}{3}}\right)/3, \\
&&  \o_{1,k}=\left(-q^*+h\,e^{-\frac{i\,2\,\pi\,k}{3}}-
q\,h^*\,e^{\frac{i\,2\,\pi\,k}{3}}\right)/3, \\
&&  \o_{2,k}=\left(-q+h^*e^{-\frac{i\,2\,\pi\,k}{3}}-
q^*h\,e^{\frac{i\,2\,\pi\,k}{3}}\right)/3.
\eea
It can be easily shown that this is the most general form for a Hamiltonian with
the local property $h_{i,i+1}^2\propto h_{i,i+1}$. The Belavin R-matrix gives
rise to Hamiltonians of the form (\ref{ekv:Hamiltonian1}) mentioned above,
\be
\label{ekv:Belavin}
R(u)=\sum_{n,m=0}^2
\,w_{nm}(u) S_{n,m}\otimes S_{2n,2m}\,.
\ee
The weights are given as
\be
w_{nm}(u)=e^{\frac{-i 2\pi}{3}(1-m)u}
\frac{\theta_1(u+\gamma+\frac{n}{3}+\frac{m\tau}{3})}
{\theta_1(\gamma+\frac{n}{3}+\frac{m\tau}{3})}\;.
\ee
This R-matrix is regular, which means that it is the permutation matrix when
$u=0$, and the Hamiltonian can then be obtained as discussed in the section two.
It also satisfies the unitarity condition, and therefore the Hamiltonian obtained
from this R-matrix will automatically satisfy Reshetikhin's condition. This condition
will now be used to show that we cannot obtain any additional integrable cases
for the Leigh-Strassler deformation.
Reshetikhin's condition simplifies when $h_{i,i+1}^2\propto h_{i,i+1}$, and
takes the form
\be
\label{ekv:enklaresh}
-2\,h_{i,i+1}h_{i+1,i+2}h_{i,i+1}+2\,h_{i+1,i+2}h_{i,i+1}h_{i+1,i+2}=
I\otimes A-A\otimes I\,.
\ee
In Appendix (\ref{Appendix1}) we show that, due to the symmetry, we will always have that if the left hand
side has a term $I\otimes A$, there is also a term $-A\otimes I$. 
The left hand term can be written explicitly as
\be
h_{i,i+1}h_{i+1,i+2}h_{i,i+1}=
\sum_{l,k,\bar{l},\bar{k}=0}^{2}\,h^L_{l\,k\,\bar{l}\,\bar{k}}\,
S_{lk}\otimes S_{\bar{l}-l, \bar{k}-k}\otimes S_{2\bar{l},2\bar{k}}
\ee
\be
h_{i+1,i+2}h_{i,i+1}h_{i+1,i+2}=\sum_{l,k,\bar{l},\bar{k}=0}^{2}\,h^R_{l\,k\,\bar{l}\,\bar{k}}\,
S_{lk}\otimes S_{\bar{l}-l, \bar{k}-k}\otimes S_{2\bar{l},2\bar{k}}\;,
\ee
where the coefficients $h^L_{l\,k\,\bar{l}\,\bar{k}}$ and $h^R_{l\,k\,\bar{l}\,\bar{k}}$ are
\bea
&& h^L_{l\,k\,\bar{l}\,\bar{k}}=
\sum_{m,n}\,\omega_{nm}\omega_{\bar{l}\bar{k}}\omega_{l-n,k-m}
\eta^{-ml+m\bar{l}-\bar{k}n+kn-\bar{k}l+\bar{l}k} \\
&& h^R_{l\,k\,\bar{l}\,\bar{k}}=
\sum_{m,n}\,\omega_{nm}\omega_{l\, k}\omega_{\bar{l}-n,\bar{k}-m}
\eta^{-m\bar{l}+m\,l-k\,n+\bar{k}n-k\,\bar{l}+l\bar{k}}
\eea
Now we would like to find all the solutions to the equation (\ref{ekv:enklaresh}). Many of these
equations are linearly dependent, so we need only a few of them.  Here we will just 
state the result, for details see Appendix (\ref{Appendix1}):
\bea
 &&h=0, \qquad q=e^{i\phi},  \\
 &&h=e^{i\,\theta},\qquad q=0\\
&&h=\rho\,e^{i\frac{2 n\pi}{3}}, \qquad q=(1+\rho)\,e^{i\,\frac{2\pi m}{3}},  \\
 &&h=e^{i\frac{2 n\pi}{3}},\qquad q=-e^{i\frac{2 m\pi}{3}} \\
&&h=0,\qquad q=0
\eea

In conclusion, we do not find any additional integrable cases. The cases $r=0$, $\rho=0$
and also $r=1$, $\rho=1$ are a bit special, in the sense that the second charge
disappears. So one might think that they are not integrable, but these are
limiting cases of Hamiltonians which have an infinite amount of commuting
charges, so this is not a problem. In another sense, one could think of them as
trivially integrable, because they are already diagonal from the beginning and
will trivially describe factorized scattering.

Now we would also like to prove that we cannot have any Hermitian
Hamiltonian of the form (\ref{ekv:Hamiltonian1}) satisfying the Reshetikhin's condition,
 except ones which can be
related to this by a $e^{i2 n\pi/3}$ site dependent shift, together with a
normal change of basis to a Hamiltonian with a $U(1)^3$ symmetry. Therefore we will
just remove the condition $h_i^2=h_i$. The coefficient $\o_{ij}$ will now be
written as
\bea
&& \o_{0,k}=\left(a+b\,e^{\frac{-i\,2\,\pi\,k}{3}}+
e^{\frac{i\,2\,\pi\,k}{3}}\right)/3\;, \\
&&  \o_{1,k}=\left(c+e\,e^{-\frac{i\,2\,\pi\,k}{3}}+
d\,e^{\frac{i\,2\,\pi\,k}{3}}\right)/3\;, \\
&&  \o_{2,k}=\left(c+e\,e^{-\frac{i\,2\,\pi\,k}{3}}+
d\,e^{\frac{i\,2\,\pi\,k}{3}}\right)/3\;.
\eea
The full Reshetikhin's condition reads
\bea
 &&h_{i,i+1}^2\,h_{i+1,i+2}-
h_{i,i+1}h_{i,i+2}^2-h_{i+1,i+2}^2h_{i,i+1}+
h_{i+1,i+2}h_{i,i+1}^2\\
&&-2\,h_{i,i+1}h_{i+1,i+2}h_{i,i+1}+2\,h_{i+1,i+2}h_{i,i+1}h_{i+1,i+2}=
I\otimes A-A\otimes I
\eea
The new terms are as follows
\be
h_{i,i+1}^2\,h_{i+1,i+2}=
\sum_{m,n}\,\omega_{nm}\omega_{\bar{l}\bar{k}}\omega_{l-n,k-m}
\eta^{ml-nk-k\bar{l}+l\bar{k}}
S_{lk}\otimes S_{\bar{l}-l, \bar{k}-k}\otimes S_{2\bar{l},2\bar{k}}
\ee
\be
h_{i,i+1}\,h_{i+1,i+2}^2=
\sum_{m,n}\,\omega_{nm}\omega_{l\,k}\omega_{\bar{l}-n,\bar{k}-m}
\eta^{m\bar{l}-n\bar{k}-k\bar{l}+l\bar{k}}
S_{lk}\otimes S_{\bar{l}-l, \bar{k}-k}\otimes S_{2\bar{l},2\bar{k}}
\ee
\be
h_{i+1,i+2}\,h_{i,i+1}^2=
\sum_{m,n}\,\omega_{nm}\omega_{\bar{l}\bar{k}}\omega_{l-n,k-m}
\eta^{ml-nk+k\bar{l}-l\bar{k}}
S_{lk}\otimes S_{\bar{l}-l, \bar{k}-k}\otimes S_{2\bar{l},2\bar{k}}
\ee
\be
h_{i+1,i+2}^2\,h_{i+1,i+2}=
\sum_{m,n}\,\omega_{nm}\omega_{l\,k}\omega_{\bar{l}-n,\bar{k}-m}
\eta^{m\bar{l}-n\bar{k}+k\bar{l}-l\bar{k}}
S_{lk}\otimes S_{\bar{l}-l, \bar{k}-k}\otimes S_{2\bar{l},2\bar{k}}
\ee
We can also show that if we have a term of the form $I\otimes A$
there is automatically a term $-A\otimes I$.
 We conclude that the only solutions
which are not immediately of a $U(1)^3$ form are the following
\be 
c_r=-\frac{d_re_r}{d_r+e_r}, \qquad b_r=1-d_r+e_r, \qquad
a_r=1-d_r-\frac{d_re_r}{d_r+e_r},
\ee
where we introduced complex polar coordinates,
as $a=a_r e^{i\phi_a}$ {\it e.t.c.}, where we allow for $a_r<0$ {\it e.t.c.}.
The only allowed phases of the complex parameters being multiples of
$2\pi/3$. 
The transformation that takes the case $h=\rho\,e^{i\frac{2 n\pi}{3}}$ 
and $q=(1+\rho)\,e^{i\,\frac{2\pi m}{3}}$ into the $q=e^{i\phi}$ case, transforms this
last case into one with a Hamiltonian where the $U(1)^3$ symmetry is apparent.

\section{Conclusion}
We have used Reshetikhin's condition to check the 
integrability properties for the dilatation
operator for both the holomorphic and the full scalar field sector of the
general Leigh-Strassler deformation to one loop. 
We have been able to  exclude the possibility of obtaining the dilatation
operator in the holomorphic sector from an R-matrix with genus one or less,
for any other  values of the parameters than the ones already found.
In the non-holomorphic sector we find that all integrable cases, except the
ones corresponding to diagonal Hamiltonians in the holomorphic sector,
stay integrable. It would be interesting to understand which factors decide 
if integrability is preserved when going from the holomorphic to the non-holomorphic sector. 
Generically this is  not  to be the case \cite{Berenstein:2004ys}.
It would be interesting to obtain the Bethe equations for the 
$h=e^{i\theta}$, $\theta \in \mathcal{R}$  with $q=0$
in the full scalar field sector. Maybe it could help in understanding
what a dual string theory background should look like. 

It is interesting to see that the simplest cases in the holomorphic sector
become so much harder in the non holomorphic sector. This should be
visible in the dual string theory as well. Even so, we have seen that the theory has even more
simple subsectors when $q=0$ and $h=0$  with non trivial eigenvalues. It would be
interesting to  study
these limiting cases from string theory. One way to start an exploration
is to consider the string background in \cite{Lunin:2005jy,Frolov:2005ty} for the $q$-deformed case, but 
the background there seems only valid for $q$ close to one.
In another article, some changes were suggested that could possibly extend the validity to arbitrary 
 $q$ values
 \cite{Chu:2006tp}.

We have found a one-loop integrable subsector to the full scalar field
sector which is integrable for any complex $q$ and $h=0$ in the closed sector
consisting of two holomorphic fields and one anti-holomorphic or vice versa.
This differs from what is the case in the holomorphic sector, where integrability
exist only for $q=e^{i\beta}$, $\beta \in \mathcal{R}$.
It is interesting to note that in this last case with a $q$-dependent factor
in front of the F-term, both integrability and the one-loop finiteness condition 
coincide. 

It would be interesting to see if this integrability also exists to higher
loops. First of all, it has to be proven whether the $\mathfrak{su}_q(2)$ sector is integrable 
or not to higher loops. If that is the case, it would be interesting to see
if an extended version of the $\mathfrak{su}_q(3)$ case is integrable to higher loops. This sector
will not be closed anymore, so we would need to include fermions and the
appropriate group would be  $\mathfrak{su}_q(2|3)$, in analogy with the non 
deformed case \cite{Beisert:2003ys,Beisert:2005fw}. Of course, before going
to higher loops it should be verified that the integrability survives the upgrading
to  $\mathfrak{su}_q(2|3)$ to one-loop. 

Another interesting question would be to look for the string Bethe equations
arising from the supergravity background suggested in \cite{Arutyunov:2004vx,Frolov:2005ty,Alday:2005ww,McLoughlin:2006cg}. 
The purpose would be to see if
the same Bethe equations 
can be derived from that background, and to check if it is the correct dual background for general complex 
parameter $q$.
It would also be interesting to look at the sigma model coming from string theory
side for this integrable sector, in the same fashion as was done in the non 
integrable holomorphic three state sector, and compare with the coherent spin 
chain sigma model obtained from this spin chain.
\acknowledgments
We would like to thank Robert Weston, Matthias Staudacher, Tristan McLoughlin and
Niklas Beisert for interesting discussions.
This work was supported by the Alexander von Humboldt foundation.
\begin{appendix}
\section{Technical details for section 5}\label{Appendix1}
Reshetikhin's condition
will now be used to show that we cannot obtain any additional integrable cases
for the Leigh-Strassler deformation.
Reshetikhin's condition simplifies when $h_{i,i+1}^2\propto h_{i,i+1}$, and
takes the form
\be
\label{ekv:enklaresh2}
-2\,h_{i,i+1}h_{i+1,i+2}h_{i,i+1}+2\,h_{i+1,i+2}h_{i,i+1}h_{i+1,i+2}=
I\otimes A-A\otimes I\,.
\ee
We will see that, due to the symmetry, we will always have that if the left hand
side has a term $I\otimes A$, there is also a term $-A\otimes I$. 
The left hand terms can be written explicitly as
\be
h_{i,i+1}h_{i+1,i+2}h_{i,i+1}=
\sum_{l,k,\bar{l},\bar{k}=0}^{2}\,h^L_{l\,k\,\bar{l}\,\bar{k}}\,
S_{lk}\otimes S_{\bar{l}-l, \bar{k}-k}\otimes S_{2\bar{l},2\bar{k}}
\ee
\be
h_{i+1,i+2}h_{i,i+1}h_{i+1,i+2}=\sum_{l,k,\bar{l},\bar{k}=0}^{2}\,h^R_{l\,k\,\bar{l}\,\bar{k}}\,
S_{lk}\otimes S_{\bar{l}-l, \bar{k}-k}\otimes S_{2\bar{l},2\bar{k}}\;,
\ee
where the coefficients $h^L_{l\,k\,\bar{l}\,\bar{k}}$ and $h^R_{l\,k\,\bar{l}\,\bar{k}}$ are
\bea
&& h^L_{l\,k\,\bar{l}\,\bar{k}}=
\sum_{m,n}\,\omega_{nm}\omega_{\bar{l}\bar{k}}\omega_{l-n,k-m}
\eta^{-ml+m\bar{l}-\bar{k}n+kn-\bar{k}l+\bar{l}k} \\
&& h^R_{l\,k\,\bar{l}\,\bar{k}}=
\sum_{m,n}\,\omega_{nm}\omega_{l\, k}\omega_{\bar{l}-n,\bar{k}-m}
\eta^{-m\bar{l}+m\,l-k\,n+\bar{k}n-k\,\bar{l}+l\bar{k}}
\eea
$S_{00}$ is the identity matrix, so the $A\otimes I$ term is
\be
\sum_{m,n}\,(\omega_{nm}\omega_{00}\omega_{l-n,k-m}
\eta^{-ml+kn}
+\omega_{nm}\omega_{l\, k}\omega_{2n,2m}
\eta^{m\,l-k\,n})
S_{lk}\otimes S_{2l, 2k}\otimes S_{0,0}\,,
\ee
and the $I\otimes A$ term is
\be
\sum_{m,n}\,(\omega_{nm}\omega_{\bar{l}\bar{k}}\omega_{l-n,k-m}
\eta^{m\bar{l}-\bar{k}n}+
\omega_{nm}\omega_{00}\omega_{\bar{l}-n,\bar{k}-m}
\eta^{-m\bar{l}+\bar{k}n})
S_{00}\otimes S_{\bar{l}, \bar{k}}\otimes S_{2\bar{l},2\bar{k}}\,.
\ee
The coefficient of $S_{00}\otimes S_{lk}\otimes S_{2l,2k}$ is equal to the
coefficient of $S_{lk}\otimes S_{2l,2k}\otimes S_{00}$.
Now we would like to find all the solutions to the equation (\ref{ekv:enklaresh}). Many of the
equations are linearly dependent, so we need only a few of them. 
We will write the $q$ and $h$ in complex polar coordinates,
as $q=-r e^{i\phi}$ and $h=\rho e^{i\theta}$, where we allow for $r,\rho<0$. The
equations coming from \{$l,k,\bar{l},\bar{k}=1,3,2,1$\} and \{$l,k,\bar{l},\bar{k}=1,3,2,2$\} 
will restrict the number of possible cases
drastically. These two equations are
\bea
&&0=\frac1{6}\rho \,r\,(e^{i\frac{2\pi}{3}}+e^{-i\frac{2\pi}{3}}\rho^2+r^2)\times \\
&&(e^{i3(\theta+\phi)}+
e^{i3(\phi)}\rho+e^{i3(\theta)}r-(e^{-i3(\theta+\phi)}+
e^{-i3(\phi)}\rho+e^{-i3(\theta)}r)(e^{i3(\theta+\phi)+i\frac{2\pi}{3}}))\nonumber
\eea
\bea
&&0=\frac1{6}\rho \,r\,(e^{i\frac{2\pi}{3}}+e^{-i\frac{2\pi}{3}}\rho^2+r^2)\times \\
&&(e^{i3(\theta+\phi)}+
e^{i3(\phi)}\rho+e^{i3(\theta)}r-(e^{-i3(\theta+\phi)}+
e^{-i3(\phi)}\rho+e^{-i3(\theta)}r)(e^{i3(\theta+\phi)-i\frac{2\pi}{3}}))\nonumber
\eea
Obviously, these two equations are satisfied when either $r=0$, $\rho=0$, or
$r^2=1$ and $\rho^2=1$. Now let us check for zeroes of the second parantheses.
There are two possibilities. Firstly, $\theta=2\pi m/3$, $\phi=2\pi n/3$ together
with $r=-1-\rho$, and secondly
$$r=-\frac{\sin 3\theta}{\sin 3(\theta-\phi)} \qquad 
\rho=\frac{\sin 3\phi}{\sin 3(\theta-\phi)}.
$$
For the first case all equations is automatically solved.
In the latter case, most of the equations will vanish, but {\it e.g.} the
equation coming from \{$l,k,\bar{l},\bar{k}=3,1,2,3$\} takes the form
\bea 
\frac{\left(1-e^{i\,6\phi}\right)^2\,\left(1-e^{i\,6\phi}\right)^2}{\left(e^{i6\theta}-e^{i6\phi}\right)^4}
(e^{i4\theta}+e^{i2\phi}+e^{2i(\theta+2\phi)}) \times \\
\left( e^{i6\theta}+e^{i2(4\,\phi+\theta)}+e^{i\frac{2\pi}{3}}(e^{i6\phi}+e^{i2(\phi+4\,\theta)})
+e^{-i\frac{2\pi}{3}}(e^{i2(\phi+\theta)}+e^{i6(\phi+\theta)})\right).
\eea
This equation restricts one of the angles to be $n\pi/3$. Further, this implies
that one of $r$ or $\rho$ is zero, and the other is one, {\it e.g.} if
$\phi=\pi/3$ we have $r=0$. So these cases are not interesting.

Now we go on to investigate the case with $r^2=1$ and $\rho^2=1$. Without loss
of generality, we can restrict to $r=1$ and $\rho=1$, since we still have a
phase factor. The equation coming from \{$l,k,\bar{l},\bar{k}=3,1,2,2$\} yields
the two real equations
\bea
&&\cos (2\theta-\phi)-e^{i\frac{9\phi}{2}} \cos (\theta-\frac{\phi}{2})=0 \\
&&\cos (2\phi-\theta)-e^{-i\frac{9\theta}{2}} \cos (\phi-\frac{\theta}{2})=0 \,,
\eea
and the equation coming from \{$l,k,\bar{l},\bar{k}=3,1,2,1$\} implies that 
either 
\be
1+2\,e^{-\frac{3\phi}{2}}\cos (-\frac{2\pi}{3}+\theta-\frac{\phi}{2})=0, 
\ee
or 
\be
3+2\left(\cos (\phi+\theta)+\cos (\theta-2\phi)+\cos (\phi-2\theta)\right)=0\;.
\ee
Analysing these equations, we conclude that the only possibilities for the
angles are $\theta=2\pi n/3$ and $\phi=2\pi n/3$.

We conclude with looking in turn at the cases $\rho=0$ and $r=0$. From the
remaining equations we get the conditions
\be
r^2-1=0 \quad \mbox{or} \quad r=0 \qquad \mbox{and} \qquad
\rho^2-1=0 \quad \mbox{or}\quad \rho=0\;,
\ee
respectively.
To conclude, the following solutions exists:
\bea
 &&h=0, \qquad q=e^{i\phi},  \\
 &&h=e^{i\,\theta},\qquad q=0\\
&&h=\rho\,e^{i\frac{2 n\pi}{3}}, \qquad q=(1+\rho)\,e^{i\,\frac{2\pi m}{3}},  \\
 &&h=e^{i\frac{2 n\pi}{3}},\qquad q=-e^{i\frac{2 m\pi}{3}} \\
&&h=0,\qquad q=0
\eea
Thus, we do not find any additional integrable cases. 
\section{Limits of the Belavin R-matrix}\label{Appendix2}
It was shown a long time ago that Cherednik's trigonometric $Z_N$ R-matrix can
be obtained as a special limit of the Belavin R-matrix \cite{Belavin:1981ix}. Not the full
Cherednik R-matrix, but only when its parameters satisfy certain relations. This
corresponds to the limit $\tau \rightarrow \infty$. We will have a look at it
and see that it is only physical for the case corresponding to the ordinary
Heisenberg spin chain. That Hamiltonian can also be obtained from directly
taking the $\gamma \rightarrow 0$ limit on the generic Hamiltonian obtained from
the original Belavin R-matrix.

To see this we first write out the elements of the R-matrix, $R=R_{km}^{ln}E_{kl}\otimes E_{mn}$
explicitly
\bea\label{parameter}
R_{ll}^{ll}=a(u)&=& w_{01}(u)+w_{02}(u)+w_{00}(u)\,,\nonumber\\
R_{l,l+1}^{l,l+1}=b(u)&=& w_{01}(u) e^{i2\pi/3}+w_{02}(u)e^{-i2\pi/3}+w_{00}(u)\,,\nonumber \\
R_{l+1,l}^{l+1,l}=\bar{b}(u)&=&
w_{02}(u) e^{i2\pi/3}+w_{01}(u)e^{-i2\pi/3}+w_{00}(u)\,,\nonumber \\
R_{l,l+1}^{l+1,l}=c(u)&=& w_{11}(u)+w_{12}(u)+w_{10}(u)\,,\nonumber\\
R_{l+1,l}^{l,l+1}=\bar{c}(u)&=& w_{21}(u)+w_{22}(u)+w_{20}(u)\,,\\
R^{l,l-1}_{l+1,l+1}=d(u)&=& w_{11}(u) e^{i2\pi/3}+w_{12}(u)e^{-i2\pi/3}+w_{10}(u)\,, \nonumber\\
R_{l,l-1}^{l+1,l+1}=\bar{d}(u)&=& w_{21}(u) e^{i2\pi/3}+w_{22}(u)
e^{-i2\pi/3}+w_{20}(u)\,, \nonumber \\
R^{l-1,l-1}_{l,l+1}=e(u)&=& w_{12}(u) e^{i2\pi/3}+w_{11}(u)e^{-i2\pi/3}+w_{10}(u)\,, \nonumber \\
R^{l,l+1}_{l-1,l-1}=\bar{e}(u)&=& w_{22}(u)
e^{i2\pi/3}+w_{21}(u)e^{-i2\pi/3}+w_{20}(u)\,, \nonumber
\eea
where
\be
w_{mn}=e^{\frac{i 2\pi\,m\,u}{3}}
\frac{\theta_1(u+\gamma+\frac{n}{3}+\frac{m\tau}{3})}
{\theta_1(\gamma+\frac{n}{3}+\frac{m\tau}{3})}
\ee
The Hamiltonian is obtained as in (\ref{forsta laddning}), expressed in terms of the
derivatives of the $\omega_{ij}$
\begin{equation}
 w_{mn}'=\frac{i 2\pi\,m}{3}+
\frac{\theta_1'(\gamma+\frac{n}{3}+\frac{m\tau}{3})}
{\theta_1(\gamma+\frac{n}{3}+\frac{m\tau}{3})}
\end{equation}
Here we see directly that the limiting case $\gamma\rightarrow 0$ corresponds to
a Heisenberg spin chain, due to the fact that the $w_{00}'$ term blows up,
leaving the Hamiltonian with just the pieces containing that term. Before we
proceed to show how the Belavin R-matrix changes shape into Cherednik's
R-matrix, we will discuss the effect of the $Z_3\times Z_3$ symmetries.

The solution is invariant under any shift
$w_{mn}\rightarrow w_{m+k_1,n+k_2}$. These shifts can be generated
from modular transformations of the above solution, taking combinations of
$\tau\rightarrow -1/\tau$ and $\tau\rightarrow \tau+1$. This is shown below. The
different transformations taking the $h=0$ case to the $q=0$ case correspond to
shifting $w_{mn}\rightarrow w_{m+2,n}$, and
$w_{mn}\rightarrow w_{nm}$ corresponds to yet another of the
transformations mentioned in \cite{Bundzik:2005zg}. The modular transformation of the
theta function is
\be
\theta_1\left(v/\tau|-1/\tau\right)=
\frac1{i}\sqrt{\frac{\tau}{i}}e^{i\pi v^2/\tau}\theta_1(v|\tau)
\ee
Using this we can rewrite $w_{mn}$ as
\bea
&& w_{mn} =-\frac1{\tau}e^{(\frac{i \pi\,m\,u}{3})}
e^{-i 2\pi (u^2+2u(\gamma+\frac{n}{3}+\frac{m\tau}{3}))/\tau}
\frac{\theta_1(u/\tau+\gamma/\tau+\frac{n}{3\tau}+\frac{m}{3}|-1/\tau)}
{\theta_1(\gamma/\tau+\frac{n}{3\tau}+\frac{m}{3}|-1/\tau)}\\
&&=-e^{-i 2\pi (u^2+2u\gamma)/\tau}\frac1{\tau}e^{(\frac{-i \pi\,n\,u/\tau}{3})}
\frac{\theta_1(u/\tau+\gamma/\tau+\frac{n}{3\tau}+\frac{m}{3}|-1/\tau)}
{\theta_1(\gamma/\tau+\frac{n}{3\tau}+\frac{m}{3}|-1/\tau)}
\eea
The first factor is an overall factor independent of $n$ and $m$, so it has no
physical implications for the Hamiltonian and can be normalized away. If we now
define $\hat{\tau}=-1/\tau$, $\hat{u}=-u/\tau$, $\hat{\gamma}=-\gamma/\tau$, the
above expression takes the shape
\be
w_{mn}\propto\frac1{\tau}e^{(\frac{i \pi\,n\,\hat{u}}{3})}
\frac{\theta_1(-\hat{u}-\hat{\gamma}-\frac{n\hat{\tau}}{3}+\frac{m}{3}|\hat{\tau})}
{\theta_1(-\hat{\gamma}-\frac{n\hat{\tau}}{3}+\frac{m}{3}|\hat{\tau})}.
\ee
We therefore conclude that the shift $\tau\rightarrow -1/\tau$ corresponds to $\omega_{nm}\rightarrow\omega_{2m,n}$.
The modular shift, $\tau\rightarrow \tau+1$, does not alter
$\theta_1(x|\tau+1)\propto \theta_1(x|\tau)$. This means that it
corresponds to an effective transformation
$w_{mn}\rightarrow w_{m,n+1}$. All changes of basis in \cite{Bundzik:2005zg}
correspond to linear combinations of these types of modular transformations.
E.g. $w_{mn}\rightarrow w_{m+2,n}$ corresponds to the three
consecutive transformations 
$\tau\rightarrow -1/\tau$, $\tau\rightarrow \tau+1$, $\tau\rightarrow -1/\tau$,
or equivalently $\tau\rightarrow \tau/(1-\tau)$. 

In this limit we have
\be
\frac{\theta_1(u+\gamma+\frac{n}{3})}
{\theta_1(\pi\gamma+\frac{n\pi}{3})}\rightarrow
\frac{\sin(\pi u+\pi\gamma+\frac{n \pi}{3})}
{\sin(\gamma+\frac{n}{3})}
\ee
and
\be
\frac{\theta_1(u+\gamma+\frac{n}{3}\pm\frac{\tau}{6})}
{\theta_1(\pi\gamma+\frac{n\pi}{3}\pm\frac{\tau}{6})}\rightarrow
 e^{\mp \pi u}
\ee
This is based on the assumption that $\gamma$ is real (taking the limit with a
non-zero imaginary part gives a diagonal matrix). In this limit, the elements
become
\bea
a(u)&=&\sin \pi u \,\cot\, 3\pi\gamma +\cos \pi u\,,\nonumber\\
b(u)&=&\sin \pi u \frac{e^{i\pi\gamma}}{\sin 3\pi \gamma} \,,\nonumber \\
\bar{b}(u)&=&\sin \pi u \frac{e^{i \pi\gamma}}{\sin 3\pi \gamma}
\nonumber \\
c(u)&=& e^{\frac{i 2\pi u}{3}} e^{-i \pi u}\,,\nonumber\\
\bar{c}(u)&=& e^{\frac{-i 2\pi u}{3}} e^{i\pi u}\,,\\
d(u)&=&0\,, \nonumber\\
\bar{d}(u)&=&0\,, \nonumber \\
e(u)&=&0\,, \nonumber \\
\bar{e}(u)&=&0\,, \nonumber \, .
\eea
We can rewrite things a little bit (redefining $i u\pi$ to  $u$ and $i\pi \gamma$ to $\gamma$ and omitting
an overall factor)
\bea
a(u)&=&1  \,,\nonumber\\
b(u)&=&\frac{\sinh u}{\sinh (u+3\gamma)} e^{\gamma} \,,\nonumber \\
\bar{b}(u)&=&\frac{\sinh u}{\sinh (u+3\gamma)} e^{-\gamma}\,,
\nonumber \\
c(u)&=& e^{\frac{u}{3}} \frac{\sinh \gamma}{\sinh (u+\gamma)}\,,\nonumber\\
\bar{c}(u)&=& e^{\frac{- u}{3}} \frac{\sinh \gamma}{\sinh (u+\gamma)}\,.
\eea
The difference now between the R-matrix above and Cherednik's R-matrix is that
in the latter the exponentials in $b$ and $\bar{b}$ can be arbitrary, $e^{-g(\gamma)}$
and $e^{g(\gamma)}$ with $g$ any function. The restriction of the function to
be $g(\gamma)=\gamma$ makes the Hamiltonian obtained from the R-matrix only Hermitian
in the limiting case $\gamma\rightarrow 0$.
Either the limit $\gamma\rightarrow 0$ can be
taken before extracting the Hamiltonian, but then we also need to take the limit
$u$ goes to zero, or the limit can be taken after the Hamiltonian has been
extracted. The first way will result in the XXX R-matrix. But both methods will in
the end result in the same Heisenberg spin chain Hamiltonian.
Using the modular transformation of the original Belavin R-matrix, we can obtain the
cases listed below:

\begin{tabular}{lll}
 $h=0$, & $q=e^{i2\,n\pi/3}$,  & $\tau\rightarrow i\infty+n$\\
 $h=e^{i\,2\,n\pi/3}$, & $q=0$, & $(\tau+n)/(1-\tau-n)\rightarrow i\infty$\\
\end{tabular}

\end{appendix}
\bibliographystyle{JHEP}
\bibliography{leighref2}

\providecommand{\href}[2]{#2}\begingroup\raggedright\begin{thebibliography}{10}

\bibitem{Maldacena:1998re}
J.~M. Maldacena, {\it {The large N limit of superconformal field theories and
  supergravity}},  {\em Adv. Theor. Math. Phys.} {\bf 2} (1998) 231--252,
  [\href{http://xxx.lanl.gov/abs/hep-th/9711200}{{\tt hep-th/9711200}}].

\bibitem{Gubser:1998bc}
S.~S. Gubser, I.~R. Klebanov, and A.~M. Polyakov, {\it Gauge theory correlators
  from non-critical string theory},  {\em Phys. Lett.} {\bf B428} (1998)
  105--114, [\href{http://xxx.lanl.gov/abs/hep-th/9802109}{{\tt
  hep-th/9802109}}].

\bibitem{Witten:1998qj}
E.~Witten, {\it {Anti-de Sitter space and holography}},  {\em Adv. Theor. Math.
  Phys.} {\bf 2} (1998) 253--291,
  [\href{http://xxx.lanl.gov/abs/hep-th/9802150}{{\tt hep-th/9802150}}].

\bibitem{Minahan:2002ve}
J.~A. Minahan and K.~Zarembo, {\it {The Bethe-ansatz for
  {$\mathcal{N}=\mathord{}$4} super Yang-Mills}},  {\em JHEP} {\bf 0303} (2003)
  013, [\href{http://xxx.lanl.gov/abs/hep-th/0212208}{{\tt hep-th/0212208}}].

\bibitem{Beisert:2003yb}
N.~Beisert and M.~Staudacher, {\it {The N = 4 SYM integrable super spin
  chain}},  {\em Nucl. Phys.} {\bf B670} (2003) 439--463,
  [\href{http://xxx.lanl.gov/abs/hep-th/0307042}{{\tt hep-th/0307042}}].

\bibitem{Beisert:2003tq}
N.~Beisert, C.~Kristjansen, and M.~Staudacher, {\it The dilatation operator of
  {$\mathcal{N}=\mathord{}$4} conformal super yang-mills theory},  {\em Nucl.
  Phys.} {\bf B664} (2003) 131--184,
  [\href{http://xxx.lanl.gov/abs/hep-th/0303060}{{\tt hep-th/0303060}}].

\bibitem{Beisert:2006ez}
N.~Beisert, B.~Eden, and M.~Staudacher, {\it Transcendentality and crossing},
  \href{http://xxx.lanl.gov/abs/hep-th/0610251}{{\tt hep-th/0610251}}.

\bibitem{Beisert:2006ib}
N.~Beisert, R.~Hernandez, and E.~Lopez, {\it A crossing-symmetric phase for
  ads(5) x s**5 strings},  {\em JHEP} {\bf 11} (2006) 070,
  [\href{http://xxx.lanl.gov/abs/hep-th/0609044}{{\tt hep-th/0609044}}].

\bibitem{Leigh:1995ep}
R.~G. Leigh and M.~J. Strassler, {\it Exactly marginal operators and duality in
  four-dimensional n=1 supersymmetric gauge theory},  {\em Nucl. Phys.} {\bf
  B447} (1995) 95--136, [\href{http://xxx.lanl.gov/abs/hep-th/9503121}{{\tt
  hep-th/9503121}}].

\bibitem{Ananth:2006ac}
S.~Ananth, S.~Kovacs, and H.~Shimada, {\it Proof of all-order finiteness for
  planar beta-deformed yang-mills},  {\em JHEP} {\bf 01} (2007) 046,
  [\href{http://xxx.lanl.gov/abs/hep-th/0609149}{{\tt hep-th/0609149}}].

\bibitem{Rossi:2006mu}
G.~C. Rossi, E.~Sokatchev, and Y.~S. Stanev, {\it On the all-order perturbative
  finiteness of the deformed n = 4 sym theory},  {\em Nucl. Phys.} {\bf B754}
  (2006) 329--350, [\href{http://xxx.lanl.gov/abs/hep-th/0606284}{{\tt
  hep-th/0606284}}].

\bibitem{Elmetti:2006gr}
F.~Elmetti, A.~Mauri, S.~Penati, and A.~Santambrogio, {\it Conformal invariance
  of the planar beta-deformed n = 4 sym theory requires beta real},  {\em JHEP}
  {\bf 01} (2007) 026, [\href{http://xxx.lanl.gov/abs/hep-th/0606125}{{\tt
  hep-th/0606125}}].

\bibitem{Lunin:2005jy}
O.~Lunin and J.~Maldacena, {\it Deforming field theories with u(1) x u(1)
  global symmetry and their gravity duals},  {\em JHEP} {\bf 05} (2005) 033,
  [\href{http://xxx.lanl.gov/abs/hep-th/0502086}{{\tt hep-th/0502086}}].

\bibitem{Frolov:2005dj}
S.~Frolov, {\it {Lax pair for strings in Lunin-Maldacena background}},
  \href{http://xxx.lanl.gov/abs/hep-th/0503201}{{\tt hep-th/0503201}}.

\bibitem{Aharony:2002hx}
O.~Aharony, B.~Kol, and S.~Yankielowicz, {\it On exactly marginal deformations
  of n = 4 sym and type iib supergravity on ads(5) x s**5},  {\em JHEP} {\bf
  06} (2002) 039, [\href{http://xxx.lanl.gov/abs/hep-th/0205090}{{\tt
  hep-th/0205090}}].

\bibitem{Fayyazuddin:2002vh}
A.~Fayyazuddin and S.~Mukhopadhyay, {\it Marginal perturbations of n = 4
  yang-mills as deformations of ads(5) x s(5)},
  \href{http://xxx.lanl.gov/abs/hep-th/0204056}{{\tt hep-th/0204056}}.

\bibitem{Kulaxizi:2006zc}
M.~Kulaxizi, {\it Marginal deformations of n = 4 sym and open vs. closed string
  parameters},  \href{http://xxx.lanl.gov/abs/hep-th/0612160}{{\tt
  hep-th/0612160}}.

\bibitem{Niarchos:2002fc}
V.~Niarchos and N.~Prezas, {\it Bmn operators for n = 1 superconformal
  yang-mills theories and associated string backgrounds},  {\em JHEP} {\bf 06}
  (2003) 015, [\href{http://xxx.lanl.gov/abs/hep-th/0212111}{{\tt
  hep-th/0212111}}].

\bibitem{Frolov:2005ty}
S.~A. Frolov, R.~Roiban, and A.~A. Tseytlin, {\it {Gauge - string duality for
  superconformal deformations of {$\mathcal{N}=\mathord{}$4} super Yang-Mills
  theory}},  \href{http://xxx.lanl.gov/abs/hep-th/0503192}{{\tt
  hep-th/0503192}}.

\bibitem{Frolov:2005iq}
S.~A. Frolov, R.~Roiban, and A.~A. Tseytlin, {\it {Gauge-string duality for
  (non)supersymmetric deformations of N = 4 super Yang-Mills theory}},  {\em
  JHEP} {\bf 07} (2005) 045,
  [\href{http://xxx.lanl.gov/abs/hep-th/0507021}{{\tt hep-th/0507021}}].

\bibitem{Kuzenko:2005gy}
S.~M. Kuzenko and A.~A. Tseytlin, {\it Effective action of beta-deformed n = 4
  sym theory and ads/cft},  \href{http://xxx.lanl.gov/abs/hep-th/0508098}{{\tt
  hep-th/0508098}}.

\bibitem{Beisert:2005if}
N.~Beisert and R.~Roiban, {\it {Beauty and the twist: The Bethe ansatz for
  twisted N = 4 SYM}},  {\em JHEP} {\bf 08} (2005) 039,
  [\href{http://xxx.lanl.gov/abs/hep-th/0505187}{{\tt hep-th/0505187}}].

\bibitem{Berenstein:2004ys}
D.~Berenstein and S.~A. Cherkis, {\it {Deformations of N = 4 SYM and integrable
  spin chain models}},  {\em Nucl. Phys.} {\bf B702} (2004) 49--85,
  [\href{http://xxx.lanl.gov/abs/hep-th/0405215}{{\tt hep-th/0405215}}].

\bibitem{Freyhult:2005ws}
L.~Freyhult, C.~Kristjansen, and T.~M\aa{}nsson, {\it {Integrable spin chains
  with $U(1)^3$ symmetry and generalized Lunin-Maldacena backgrounds}},
  \href{http://xxx.lanl.gov/abs/hep-th/0510221}{{\tt hep-th/0510221}}.

\bibitem{Alday:2005ww}
L.~F. Alday, G.~Arutyunov, and S.~Frolov, {\it Green-schwarz strings in
  tst-transformed backgrounds},  {\em JHEP} {\bf 06} (2006) 018,
  [\href{http://xxx.lanl.gov/abs/hep-th/0512253}{{\tt hep-th/0512253}}].

\bibitem{Roiban:2003dw}
R.~Roiban, {\it On spin chains and field theories},  {\em JHEP} {\bf 09} (2004)
  023, [\href{http://xxx.lanl.gov/abs/hep-th/0312218}{{\tt hep-th/0312218}}].

\bibitem{Bundzik:2005zg}
D.~Bundzik and T.~M\aa{}nsson, {\it The general leigh-strassler deformation and
  integrability},  {\em JHEP} {\bf 01} (2006) 116,
  [\href{http://xxx.lanl.gov/abs/hep-th/0512093}{{\tt hep-th/0512093}}].

\bibitem{DeVega:1988ry}
H.~J. De~Vega, {\it Yang-baxter algebras, integrable theories and quantum
  groups},  {\em Int. J. Mod. Phys.} {\bf A4} (1989) 2371--2463.

\bibitem{McLoughlin:2006cg}
T.~McLoughlin and I.~Swanson, {\it Integrable twists in ads/cft},  {\em JHEP}
  {\bf 08} (2006) 084, [\href{http://xxx.lanl.gov/abs/hep-th/0605018}{{\tt
  hep-th/0605018}}].

\bibitem{Beisert:2005wv}
N.~Beisert and T.~Klose, {\it Long-range gl(n) integrable spin chains and
  plane-wave matrix theory},  {\em J. Stat. Mech.} {\bf 0607} (2006) P006,
  [\href{http://xxx.lanl.gov/abs/hep-th/0510124}{{\tt hep-th/0510124}}].

\bibitem{Belavin:1981ix}
A.~A. Belavin, {\it Dynamical symmetry of integrable quantum systems},  {\em
  Nucl. Phys.} {\bf B180} (1981) 189--200.

\bibitem{Gomez:1996az}
C.~Gomez, G.~Sierra, and M.~Ruiz-Altaba, {\it Quantum groups in two-dimensional
  physics}, . Cambridge, UK: Univ. Pr. (1996) 457 p.

\bibitem{Tetelman:1982}
M.~G. Tetelman {\em Sov.Phys. JETP} {\bf 55(2)} (1982) 306--310.

\bibitem{Madhu:2007ew}
K.~Madhu and S.~Govindarajan, {\it Chiral primaries in the leigh-strassler
  deformed n = 4 sym: A perturbative study},
  \href{http://xxx.lanl.gov/abs/hep-th/0703020}{{\tt hep-th/0703020}}.

\bibitem{Freedman:2005cg}
D.~Z. Freedman and U.~Gursoy, {\it Comments on the beta-deformed n = 4 sym
  theory},  {\em JHEP} {\bf 11} (2005) 042,
  [\href{http://xxx.lanl.gov/abs/hep-th/0506128}{{\tt hep-th/0506128}}].

\bibitem{Parkes:1984dh}
A.~Parkes and P.~C. West, {\it Finiteness in rigid supersymmetric theories},
  {\em Phys. Lett.} {\bf B138} (1984) 99.

\bibitem{Jones:1984cx}
D.~R.~T. Jones and L.~Mezincescu, {\it The chiral anomaly and a class of two
  loop finite supersymmetric gauge theories},  {\em Phys. Lett.} {\bf B138}
  (1984) 293.

\bibitem{Chu:2006tp}
C.-S. Chu and V.~V. Khoze, {\it String theory dual of the beta-deformed gauge
  theory},  \href{http://xxx.lanl.gov/abs/hep-th/0603207}{{\tt
  hep-th/0603207}}.

\bibitem{Beisert:2003ys}
N.~Beisert, {\it The su(2|3) dynamic spin chain},  {\em Nucl. Phys.} {\bf B682}
  (2004) 487--520, [\href{http://xxx.lanl.gov/abs/hep-th/0310252}{{\tt
  hep-th/0310252}}].

\bibitem{Beisert:2005fw}
N.~Beisert and M.~Staudacher, {\it Long-range psu(2,2|4) bethe ansaetze for
  gauge theory and strings},  {\em Nucl. Phys.} {\bf B727} (2005) 1--62,
  [\href{http://xxx.lanl.gov/abs/hep-th/0504190}{{\tt hep-th/0504190}}].

\bibitem{Arutyunov:2004vx}
G.~Arutyunov, S.~Frolov, and M.~Staudacher, {\it Bethe ansatz for quantum
  strings},  {\em JHEP} {\bf 10} (2004) 016,
  [\href{http://xxx.lanl.gov/abs/hep-th/0406256}{{\tt hep-th/0406256}}].

\end{thebibliography}\endgroup

\end{document}